\documentclass[a4paper,11pt]{article}
\usepackage{jheppub}
\usepackage[colorlinks=true, linkcolor=blue, citecolor=blue, urlcolor=blue]{hyperref}
\usepackage{graphicx, multirow, chngpage}
\usepackage{amsmath,amssymb,amsfonts,physics,slashed}
\usepackage{caption,subcaption}
\usepackage[T1]{fontenc}
\usepackage{bbold}
\usepackage{booktabs}
\usepackage[makeroom]{cancel} 

\usepackage{xcolor}
\usepackage[normalem]{ulem}

\renewcommand{\vec}[1]{{\boldsymbol{#1}}} 

\title{A Chiral Nelson-Barr Relaxion}

\author{Cristien Arzate}
\author{and Joshua Iascau}
\affiliation{Department of Physics and Institute for Fundamental Science, \\University of Oregon, Eugene, Oregon 97403, USA}

\emailAdd{cristien@uoregon.edu, joshuai@uoregon.edu}

\abstract{We present the chiral Nelson–Barr relaxion model, an extension of the Nelson–Barr relaxion framework with a gauged chiral $\mathrm{U}(1)$ symmetry. It includes vector-like quarks to cancel gauge anomalies and a scalar whose vacuum expectation value generates their masses. The chiral symmetry forbids the leading operators responsible for the Nelson-Barr and clockwork quality problems. Right-handed neutrinos generate light-neutrino masses through the type-I seesaw mechanism and permit thermal leptogenesis. For the particular Nelson-Barr sector Yukawa coupling considered, relaxion phenomenology gives a maximal scale of $CP$ violation $f_{\textrm{max}}^{\textrm{pheno}}\sim 4 \times 10^{11}\ \mathrm{GeV}$, about an order of magnitude greater than the corresponding phenomenological ceiling in the original model. The model can accommodate thermal leptogenesis if its right-handed-neutrino masses and reheating history satisfy the Davidson-Ibarra bound.}


\begin{document}
\maketitle
\flushbottom

\section{Introduction} 
\label{sec:intro}

The Standard Model (SM) leaves at least two naturalness puzzles unaddressed: the electroweak (EW) hierarchy problem \cite{giudice_2008, peskin2025hierarchyproblem} and the strong $CP$ problem \cite{tHooft:1976rip, hook2023tasilecturesstrongcp, ai2026statusstrongcpproblem}. The EW hierarchy problem asks why the Higgs mass sits orders of magnitude below the scale at which new physics is expected to appear, while the strong $CP$ problem asks why the angle $\bar\theta$ (a sum of the topological QCD vacuum angle and the phase of the determinant of the up and down quark mass matrices), which is a priori an $\mathcal{O}(1)$ parameter, is observationally constrained to be smaller than $10^{-10}$ \cite{PhysRevLett.124.081803}. Various solutions targeting both problems have been proposed \cite{davidi_hierarchion_2018, PhysRevD.100.055003}. However, such models contain quality problems caused by higher-dimensional operators. We present a model which addresses both the electroweak hierarchy and strong $CP$ problems while resolving these quality issues. The model also addresses neutrino masses and baryogenesis. We call it the chiral Nelson-Barr relaxion ($\chi$NBR) model.

Cosmological relaxation of the electroweak scale \cite{PhysRevLett.115.221801} offers a nice alternative to other solutions of the hierarchy problem: a slowly rolling scalar field, the relaxion, scans the Higgs mass during inflation and dynamically comes to rest at the observed value. In its original incarnation, the relaxion can be identified with the QCD axion, so that the same field responsible for stabilizing the electroweak scale also solves the strong CP problem via the Peccei-Quinn mechanism. However, models of this type suffer from a "relaxion $CP$ problem" \cite{PhysRevD.99.035014}, as they do not naturally yield a small enough $\bar{\theta}$.

The Nelson-Barr (NB) mechanism \cite{PhysRevLett.53.329, NELSON1984387, NELSON1984165}, later cast into its minimal form by Bento, Branco, and Parada (BBP) \cite{Bento:1991ez}, offers a distinct route to a naturally small $\bar\theta$: $CP$ is taken to be an exact symmetry of the UV theory, spontaneously broken by the vacuum expectation value (VEV) of a new scalar, with the resulting phase communicated to the SM Cabibbo--Kobayashi--Maskawa (CKM) matrix through mixing with a heavy vector-like quark (VLQ), while $\bar\theta$ itself is protected from receiving a large contribution at tree level by the specific structure of the new scalar and VLQ couplings. Combining this NB sector with a relaxion sector solves the relaxion $CP$ problem outlined above, giving rise to the NB relaxion (NBR) of \cite{PhysRevD.99.035014}, which we review in Sec.~\ref{sec:nbrelaxion}.

A third ingredient is needed to complete the NBR construction. Successfully combining the relaxion and NB sectors requires a large hierarchy between the periodicity of the relaxion's "roll" potential and that of its "backreaction" potential. The clockwork mechanism \cite{PhysRevD.90.023545, Choi_2016, PhysRevD.93.085007} supplies exactly this kind of hierarchy: a chain of $N+1$ scalars with nearest-neighbor couplings which, after symmetry breaking, yield a single light mode whose $N$-th site coupling is exponentially suppressed in $N$ without the need to tune any input parameter to be small. We review the clockwork construction and its embedding into the NBR framework in Sec.~\ref{sec:clockwork-intro}.

Global symmetries of the kind employed by the clockwork sector are expected to be broken by Planck-suppressed operators in any consistent theory of quantum gravity \cite{harlow_symmetries_2021, reece2023tasilecturesnoglobal}. Left unaddressed, such operators reintroduce large, uncontrolled contributions to the relaxion potential, spoiling the clockwork mechanism. This is something we call the "clockwork quality problem" \cite{higaki_quality_2016} (the authors of \cite{PhysRevD.99.035014} call it the "relaxion quality problem"). In any effective field theory (EFT), there will exist EFT-scale-suppressed higher-dimensional non-renormalizable operators which respect the symmetries of the IR, which means that, in the NB sector, there are potential contributions to $\bar{\theta}$ which could be ruinous for NB constructions. This is known as the "NB quality problem" \cite{Choi_1993, dine_challenges_2015}. As we discuss in Sec.~\ref{sec:quality}, the NBR model of \cite{PhysRevD.99.035014} inherits both quality problems simultaneously, and satisfying the NB quality bound restricts the scale of $CP$ violation to $f \lesssim 10^8$ GeV. As we show, this is in direct tension with the Davidson--Ibarra (DI) bound \cite{DAVIDSON200225} of thermal leptogenesis models, which, for standard hierarchical thermal leptogenesis, implies a reheating lower bound. Together with the domain-wall requirement $T_{\rm reh} \lesssim f$ and the NB-quality ceiling, this leaves no robust allowed interval in the minimal NBR setup.

A resolution to the NB quality problem was recently proposed by Asadi, Homiller, Lu, and Reece \cite{PhysRevD.107.115012}, who showed that gauging a chiral $\mathrm{U}(1)$ symmetry can forbid the dangerous operators responsible for the quality problem, provided the resulting gauge anomalies are cancelled by an additional VLQ. In this paper, we apply this strategy to the NBR of \cite{PhysRevD.99.035014}, extending the gauged $\mathrm{U}(1)$ to act on the clockwork sector (and the relaxion sector) as well as the NB sector. We show that a single chiral gauge symmetry, appropriately charged across both sectors, simultaneously resolves the clockwork and NB quality problems and raises the NB-quality ceiling to $f \lesssim 10^{13}$ GeV. Separately, for the benchmark couplings studied here, relaxion phenomenology gives $f_{\max}^{\rm pheno}\sim 4 \times 10^{11}\ \mathrm{GeV}$, about eight times the corresponding phenomenological ceiling in the original NBR model. With right-handed neutrinos, the parameter space can accommodate thermal leptogenesis, subject to the reheating-temperature requirement.

The rest of this paper is organized as follows:\\

\noindent \textbf{Review (Sections~\ref{sec:nbrelaxion}-~\ref{sec:quality}):}
\begin{itemize}
    \item In Sec.~\ref{sec:nbrelaxion}, we review the Nelson-Barr relaxion (NBR) model of \cite{PhysRevD.99.035014}, including:
    \begin{itemize}
        \item the clockwork mechanism (Sec.~\ref{sec:clockwork-intro})
        \item the NB sector (Sec.~\ref{sec:nb_sector})
        \item the one-loop generation of the relaxion's rolling potential (Sec.~\ref{sec:1lnb_to_roll}) 
        \item radiative contributions to $\bar{\theta}$ (Sec.~\ref{sec:radcorr_nbr})
    \end{itemize}
    \item In Sec.~\ref{sec:quality}, we discuss the clockwork and NB quality problems and their tension with thermal leptogenesis
\end{itemize}

\noindent \textbf{Novel work (Sections~\ref{sec:chiral_nb_relaxion}-~\ref{sec:pheno}):}
\begin{itemize}
    \item In Sec.~\ref{sec:chiral_nb_relaxion}, we construct the $\chi$NBR model and:
    \begin{itemize}
        \item derive constraints on the NB Yukawa coupling (Sec.~\ref{sec:nbyukcons})
        \item identify an anomaly-free $\mathrm{U}(1)_\chi$ charge assignment (Sec.~\ref{sec:anom_canc}) 
        \item demonstrate how our charge assignment resolves both quality problems (Sec.~\ref{sec:on_cw_nb_qp})
    \end{itemize}
    \item In Sec.~\ref{sec:pheno}, we present the phenomenology of the $\chi$NBR model and explore its parameter space
\end{itemize}

We conclude in Sec.~\ref{sec:conc}. The appendices provide details of the mass diagonalization, clockwork construction, radiative roll potential, and relaxion--Higgs mixing calculation.

\section{The Nelson-Barr relaxion}
\label{sec:nbrelaxion}

The NBR model \cite{PhysRevD.99.035014} is composed of three sectors in addition to the SM, such that its Lagrangian is described by
\begin{equation}
    \mathcal{L}_{\textrm{NBR}} = \mathcal{L}_{\textrm{SM}} + \mathcal{L}_{\textrm{R}} + \mathcal{L}_{\textrm{NB}} + \mathcal{L}_{\textrm{clock}}
\end{equation}
The Lagrangian $\mathcal{L}_{\textrm{R}}$ is based on the compact relaxion model found in \cite{Flacke_2017}, which in turn was a modification of the relaxion models in \cite{Gupta_2016}, which in yet another turn was a change to the original model of Graham, Kaplan and Rajendran \cite{PhysRevLett.115.221801}. The following relaxion sector identified as $\mathcal{L}_{\textrm{R}}$ can be written in terms of two different potentials: $\mathcal{L}_{\textrm{R}} = -V_{\textrm{roll}} - V_{\textrm{br}}$. $V_{\textrm{roll}}$ is known as the roll potential and $V_{\textrm{br}}$ is known as the backreaction potential. The roll potential is
\begin{equation}\label{eq:vroll}
V_{\textrm{roll}}(\phi, h) = \qty(\kappa \Lambda_{H}^2 - \Lambda_H^2 \cos(\frac{\phi}{F}) ) H^\dagger H + \lambda (H^\dagger H)^2 - r_{\textrm{roll}}^2 \Lambda_{H}^4 \cos(\frac{\phi}{F}),
\end{equation}
while the backreaction potential is equal to
\begin{equation}\label{eq:vbr}
V_{\textrm{br}}(\phi, h) = -M_{\textrm{br}}^2 H^{\dagger} H \cos(\frac{\phi}{f}) - r_{\textrm{br}}^2 M_{\textrm{br}}^4 \cos(\frac{\phi}{f}).
\end{equation}
The field $\phi$ is the relaxion, $H$ is the SM Higgs doublet, and $h$ is the SM Higgs field. The $H^{\dag} H$ term in Eq.~\ref{eq:vroll} is the Higgs mass term. Letting $\mu^2(\phi) = \kappa \Lambda_{H}^2 - \Lambda_H^2 \cos(\frac{\phi}{F})$, we see that the Higgs mass is controlled by the relaxion field. In the original relaxion setup, the last term in $V_\textrm{roll}$ comes from non-perturbative instanton effects (like an axion, the original relaxion couples to the field strength operator $G \tilde{G}$ of some strong sector and brings about this term). In this setup, however, the last term (and the other term that is proportional to $\cos(\phi/F)$) comes one-loop radiative corrections to the NB Lagrangian $\mathcal{L}_{\textrm{NB}}$. We discuss this in more detail in Sec.~\ref{sec:1lnb_to_roll}. Generation of the backreaction potential (from a UV completion) is a discussion left to the references \cite{Gupta_2016, Flacke_2017, davidi_hierarchion_2018}. $F$ ($f$) is the large (small) periodicity of the roll (backreaction) potential, $\Lambda_H$ is the UV cutoff of the Higgs effective theory in which the relaxion is embedded, and $M_{\textrm{br}}$ parametrizes the mass threshold of the backreaction scale $\Lambda_{\textrm{br}} = \sqrt{M_{\textrm{br}} v}$. Note that $\kappa \lesssim 1$ and $r_{\textrm{roll}}, r_{\textrm{br}} \gtrsim 1 / 4 \pi$. The bound on $r_{\textrm{roll}}$ can be found by calculating the loop attained by contracting the ends of two of the Higgs lines from the quartic Higgs interaction in Eq.~\ref{eq:vroll}. Finding the bound on $r_{\textrm{br}}$ takes a little more work, so we refer the reader to \cite{PhysRevLett.115.251803, Gupta_2016, Flacke_2017}.

The exact initial field value of the relaxion is not vital to this construction. As long as the initial value of $\phi$ is such that  $\mu^2(\phi) > 0$, the relaxion will roll down its potential during inflation until it reaches the critical point where $\mu^2(\phi) = 0$. At this point, EW symmetry gets broken, the Higgs develops a vacuum expectation value (VEV), and the backreaction potential generates wiggles which function to stop the evolution of the relaxion at $\phi = \ev{\phi}$ such that $\qty|\mu^2(\ev{\phi})| = m_h^2$. 

The relaxion model coupled with a model of inflation (the original proposal relies on slow-roll inflation, but see \cite{Ibe_2019} for an example of a fast-rolling relaxion) solves the hierarchy problem. There is a caveat, however. Inflation models upon which the relaxion relies require field excursions exceeding the Planck scale \cite{LINDE1983177}. For our model this means that if our backreaction potential has a periodicity $f$, then the periodicity $F$ of the roll potential, must be such that $F \gg f$. Since this large ratio of periodicities is not an inherent feature of the relaxion sector, a way to realize this separation of scales is where we turn to next.

\subsection{Clockwork Mechanism }
\label{sec:clockwork-intro}

The clockwork mechanism \cite{PhysRevD.90.023545, Choi_2016, PhysRevD.93.085007} can be viewed as the answer to the question of how to generate the hierarchy of scales $F \gg f$, so, naturally, we explain it here. The model includes $N+1$ complex scalars $\Phi_j$, where $j = 0, \dots, N$. Their Lagrangian $\mathcal{L}_{\textrm{clock}}$ includes kinetic terms for the complex scalars and a potential 
\begin{equation}
    V_{\textrm{clock}}(\Phi) = \sum_{j = 0}^N \qty( -m^2 |\Phi_j|^2 + g^2_{\textrm{clock}} |\Phi_j|^4 ) + \Delta V_{\textrm{clock}}(\Phi),
\end{equation}
where 
\begin{equation}
    \Delta V_{\textrm{clock}}(\Phi) = \sum_{j = 0}^{N - 1} \qty( \varepsilon \Phi_j^{\dag} \Phi_{j + 1}^3 + \textrm{H.c.} ).
\end{equation}
The potential $V_{\textrm{clock}}(\Phi)$ would respect a global $\mathrm{U}(1)^{N+1}$ symmetry, but $\Delta V_{\textrm{clock}}(\Phi)$ explicitly breaks this symmetry down to a single $\mathrm{U}(1)$ symmetry which we call $\mathrm{U}(1)_N$. Under the $\mathrm{U}(1)_N$ symmetry, the fields $\Phi_j$ have charges $Q = 1, \frac{1}{3}, \frac{1}{9}, \dots, \frac{1}{3^N}$.
Letting $\varepsilon \ll g^2_{\textrm{clock}} < 1$, we can take the limit $\varepsilon \rightarrow 0$ and then expand around the scalar VEVs $\abs{\langle \Phi_j \rangle} = \frac{f}{\sqrt{2}} \ \forall \ j$. Below the scale $f$ we can assume the fields are all at their VEVs such that
\begin{equation}
    \Phi_j \rightarrow U_j \equiv \frac{f}{\sqrt{2}} \mathrm{e}^{i \pi_j / f}.
\end{equation}
This yields a mass matrix $M$ for the $\pi_j$ proportional to $\epsilon f^2 / 4$.
In the case of finite $N$, the mass eigenmodes ${a^j}$ contain a zero-mode which has an exponentially suppressed overlap $\sim 3^{-N}$ with whatever operator $\Phi_N$ couples to in the UV. 

If we identify the effective decay constant $f$ with the periodicity of the backreaction potential in Eq.~\ref{eq:vbr} this is essentially the same as placing the backreaction potential at the zeroth clockwork site. In the next section we will see that we have to place the NB sector at the $N$-th site so that its loop corrections can generate the roll potential with periodicity $F$. More details on the clockwork construction can be found in Appendix~\ref{sec:det_clockmech}.

\subsection{Nelson-Barr Sector}
\label{sec:nb_sector}

The NB class of models was first developed as a way to solve the strong $CP$ problem in the context of grand unified theories \cite{PhysRevLett.53.329, NELSON1984387, NELSON1984165} and, by Bento, Branco, and Parada (BBP), was later put into its modern form as a theory built upon the SM with only an additional scalar and a vector-like quark (VLQ) (this forms a class of models known as minimal NB models) \cite{Bento:1991ez}. We will refer to the original minimal NB model as the BBP model. Like the BBP model, we introduce a new scalar $\Phi_N$ and a VLQ $U$ (where $U$ refers to a Dirac spinor in the Weyl representation composed of Weyl spinors $U_L$ and $U_R$). However, instead of using a down-type mediated NB sector, we utilize an up-type mediated one, where $\Phi_N$ and $U$ couple to the SM up quark sector through Yukawa interactions. We do so to remain consistent with the choices made in constructing the NBR model. This modified BBP model Lagrangian $\mathcal{L}_{\textrm{NB}}$ can be written in terms of a potential
\begin{equation}\label{eq:nbr}
    V_{\textrm{NB}} \supset \mu_U \bar{U}_L U_R + \qty( y^U_i \Phi_N + \tilde{y}^U_i \Phi_N^{*} ) \bar{U}_L u^i_R + (Y^u)^i_{\ j} \bar{Q}_{L,i} \tilde{H} u^j_R + \textrm{H.c.}.
\end{equation}
$\bar{U}_L$ and $U_R$ are $\textrm{SU}(2)_R$ VLQ singlets with hypercharges $\mp 2/3$. We use the two-component bar notation of \cite{wess_supersymmetry_1992}, where, for example: $\bar{U}_R^{\alpha} \equiv (U_R^{\dot{\alpha}})^\dag$. Note that except for the previous expression, we've kept spinor indices suppressed. $Q_L$ refers to the left-handed SM quark doublets and $u_R$ refers to the right-handed SM quark singlets. $H$ is the SM Higgs doublet, the $i$ are flavor indices which run from $i = 1, \dots, 3$, and the $Y^u$ are SM up quark Yukawas. $\Phi_N$ is the scalar at the $N$-th clockwork site and $y^U_i$ and $\tilde{y}^U_i$ are BSM Yukawas that couple $\Phi_N$ to $U$ and the SM right-handed quark singlets. Note that the VLQ mass $\mu_U$ and all of the couplings are real. In addition, we assume that our Lagrangian is in a basis where the down quark mass matrix is real diagonal, and assume a minimal flavor violation (MFV) scenario such that $\sum_{ij} \left[ Y^{u \dag} Y^u \right]_{i j} \approx \sum_{ij} y_t^{2} V^*_{3 i} V_{3 j} \approx 2 (\frac{m_t}{v})^{2} \cdot 10^{-2} \sim \mathcal{O}(10^{-2})$ \cite{DAMBROSIO2002155, PhysRevD.80.076002, isidori_minimal_2012}. 

\noindent The scalar $\Phi_N$ and the VLQs $\bar{U}_L$ and $U_R$ are also charged under a $\mathbb{Z}_2$ symmetry such that
\begin{equation}
    \Phi_N \rightarrow \mathrm{e}^{i \pi k} \Phi_N, \quad \bar{U}_L \rightarrow \mathrm{e}^{i \pi k} \bar{U}_L, \quad U_R \rightarrow \mathrm{e}^{-i \pi k} U_R.
\end{equation}
In addition to Eq.~\ref{eq:nbr}, the scalar sector of the NB potential includes 
\begin{equation}
    V_{\textrm{NB}} \supset \lambda |\Phi_N|^2 |H|^2.
\end{equation}
Note that in the BBP model, spurious terms of the form $\Phi_N^2 + \Phi_N^{*2}$ and $\Phi_N^4 + \Phi_N^{*4}$ are needed to produce a spontaneously $CP$-violating VEV. However, in the NBR model these terms are unnecessary, as the $CP$-violating VEV ultimately comes from relaxion dynamics. The tree-level up quark mass matrix after the new scalar and the Higgs get VEVs, which are $\ev{\Phi_N} = \frac{f}{\sqrt{2}} \mathrm{e}^{i \phi_0 / F} = \frac{f}{\sqrt{2}} \mathrm{e}^{i \theta_N}$ and $\ev{H} = \frac{v}{\sqrt{2}}$, respectively, then looks like
\begin{equation}
    V_{\textrm{NB}} \supset \pmqty{ \bar{u} & \bar{U} }_L \mathcal{M}_0 \pmqty{ u \\ U }_R, \quad \mathcal{M}_0 = \pmqty{ m_u & 0 \\ F & \mu_U }
\end{equation}
where $(m_u)^i_{\ j} = (Y^u)^i_{\ j} v / \sqrt{2}$ and $F_i = y^U_i \langle \Phi_N \rangle + \tilde{y}^U_i \langle \Phi^{*}_N \rangle$. Because $F_i$ is the only complex quantity in the mass matrix and it is diagonally across from the zero entry in the mass matrix, it doesn't contribute any complex part to the determinant of the matrix, therefore the NB Lagrangian is $CP$ symmetric at tree level. 

To find the effect that the NB sector has on the SM up quark masses, we can partially diagonalize our mass matrix using a bi-unitary transformation. The details of how exactly this is done are found in Appendix~\ref{sec:pdiag_mm}. We find that the effective up quark mass matrix is given by
\begin{equation}
    K \bar{m}^2 K^{ -1 } = \frac{v}{2} y^u y^{u \dag} \approx m_u \qty(\mathbb{1} - \frac{F^{ \dag } F}{ F F^{ \dag } + \mu_U^2 }) m_u^{\intercal } .
\end{equation}
To produce an $\mathcal{O}(1)$ phase we must enforce that $F_i F_i^{ \dag } \gtrsim \mu_U^2$. This ensures that the second term in the parentheses will be $\mathcal{O}(1)$. Since the CKM matrix must be unitary, we also enforce that $y_i^U \sim \tilde{y}_i^U \sim y^U$ for all $i$ and that $F, \mu_U \gg m_u$.

\subsection{One-loop Nelson-Barr Generation of Roll Potential}
\label{sec:1lnb_to_roll}

\begin{figure}
\begin{subfigure}{0.45\textwidth} 
\centering
\includegraphics[width=\textwidth]{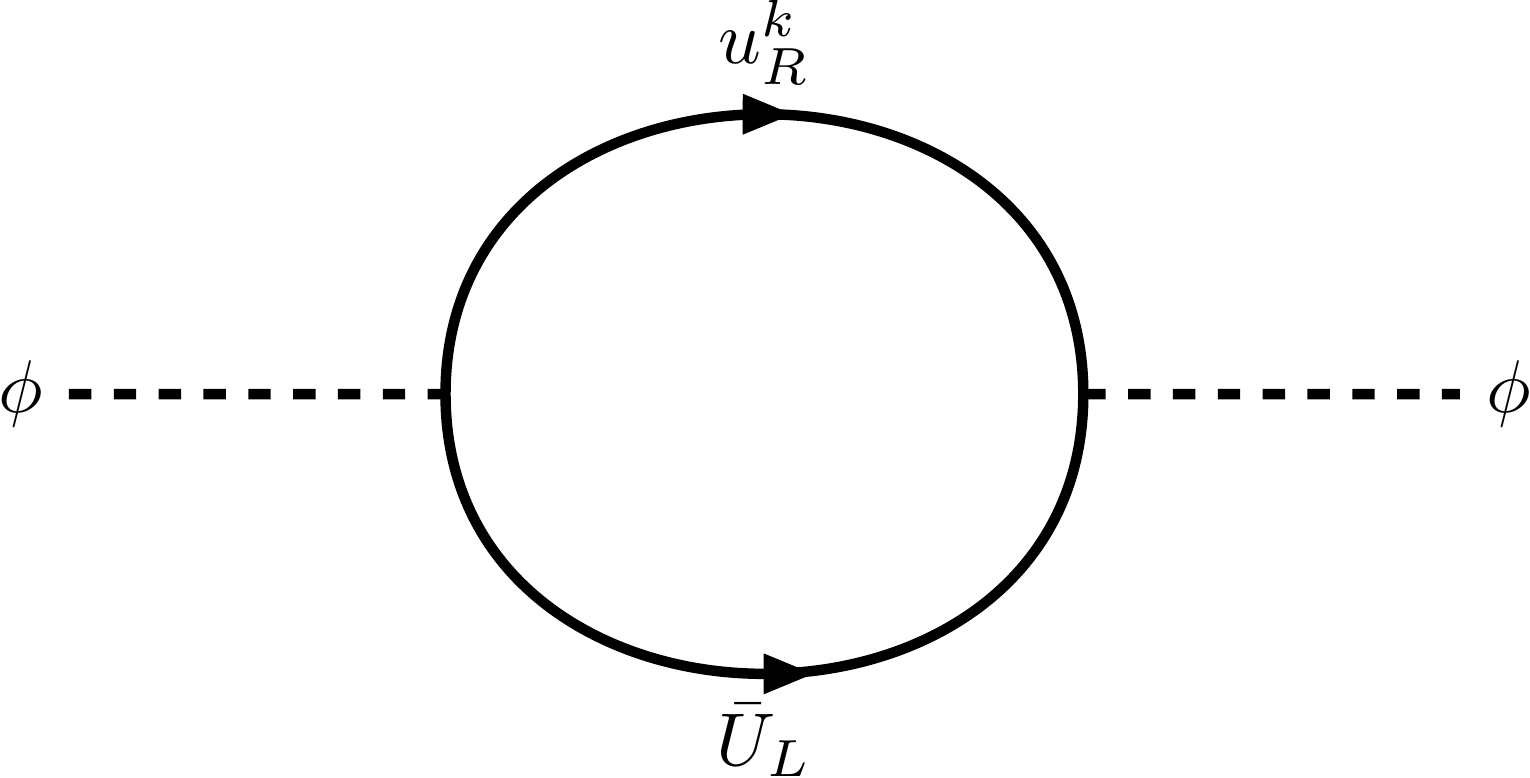}
\caption{ $\phi \rightarrow \phi$}
\label{fig:diag_phiphi}
\end{subfigure}
\hfill
\begin{subfigure}{0.45\textwidth} 
\centering
\includegraphics[width=\textwidth]{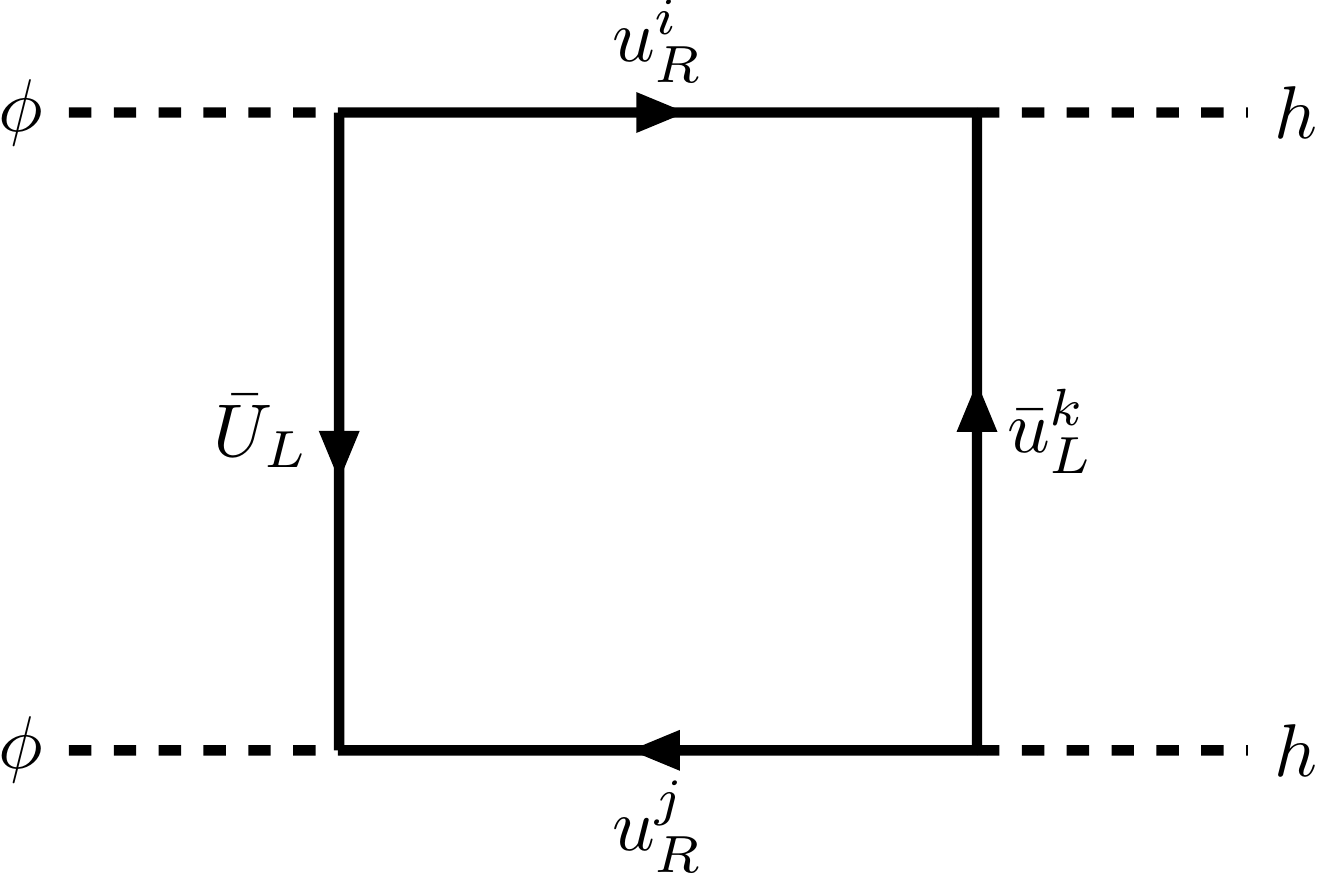}
\caption{$\phi \phi \rightarrow h h$}
\label{fig:diag_phiphihh}
\end{subfigure}
\centering
\caption{In (~\ref{fig:diag_phiphi}) is a $\phi \rightarrow \phi$ one-loop correction and in (~\ref{fig:diag_phiphihh}) is the $\phi \phi \rightarrow h h$ box diagram.}
\label{fig:feynmandiagrams}
\end{figure}

The roll potential is generated at one-loop level from the $\phi\rightarrow\phi$ loop diagram and the $\phi\phi\rightarrow h h$ quark box diagram. Both of these diagrams are found in Fig.~\ref{fig:feynmandiagrams}. 
The $\phi\rightarrow\phi$ one-loop self-energy $-i I_1$ computed with a hard-momentum cutoff at loop momentum $k=f$ is 
\begin{equation}
    -i I_1 \approx -i \frac{f^4}{16 \pi^2 F^2} y_k^{U} \tilde{y}_k^{U}.
\end{equation}
This corrects the mass such that it is $m^{2} = m_0^{2} + I_1$, where $m_0$ is the bare mass. Thus, it yields an additional term in the potential that looks like $\frac{1}{2} I_1 |\phi|^2$. 
Now, since we have,
\begin{align*}
    V_{\textrm{roll}} \supset & -\Lambda_{\textrm{roll}}^4 \cos(\frac{\phi}{F}) = -r_{\textrm{roll}}^2 \Lambda_H^4 \cos(\frac{\phi}{F}), \\
    \approx & -r_{\textrm{roll}}^2 \Lambda_H^4 \qty(1 - \frac{1}{2} \frac{|\phi|^2}{F^2}) \rightarrow \frac{1}{2} \frac{r_{\textrm{roll}}^2}{F^2} \Lambda_H^4 |\phi|^2,
\end{align*}
we must require that
\begin{equation*}
    \frac{1}{2} \frac{r_{\textrm{roll}}^2}{F^2} \Lambda_H^4 |\phi|^2 = \frac{1}{2} I_1 |\phi|^2,
\end{equation*}
which leads to
\begin{equation}\label{eq:rroll_pre}
    r_{\textrm{roll}} = \sqrt{y_k^{U} \tilde{y}_k^{U}} \frac{f^2}{4 \pi \Lambda_H^2}.
\end{equation}

To finish this calculation, we need an expression for $\Lambda_H^2$. 
For this we must compute the amplitude of the quark loop diagram for the process $\phi \phi \rightarrow h h$. The loop, call it $I_2$, is found to be
\begin{equation}
    i I_2 \approx -i \frac{1}{64 \pi^2} \frac{f^2}{F^2} y_i^{U} \tilde{y}_j^{U} \qty(Y^{u \dag} Y^u)_{ij}.
\end{equation}
Since the amplitude $I_2$ corrects the 4-point coupling to $\qty|\phi|^2 h^2$ as $\lambda' = \lambda - I_2$, matching to the roll potential yields the relation
\begin{equation}
    V_{\textrm{roll}} \supset -\Lambda_H^2 \cos(\frac{\phi}{F}) H^\dagger H \rightarrow \frac{\Lambda_H^2}{4 F^2} \qty|\phi|^2 h^2 = -I_2 \qty|\phi|^2 h^2,
\end{equation}
which leads to
\begin{equation}\label{eq:lambdah}
    \Lambda_H \approx \frac{\sqrt{y_i^{U} \tilde{y}_j^{U} \qty(Y^{u \dag} Y^u)_{ij}}}{4 \pi} f.
\end{equation}
Plugging this back into Eq.~\ref{eq:rroll_pre}, we find that
\begin{equation}\label{eq:rroll}
    r_{\textrm{roll}} \approx \frac{4 \pi \sqrt{y_k^{U} \tilde{y}_k^{U}}}{y_i^{U} \tilde{y}_j^{U} \qty(Y^{u \dag} Y^u)_{ij}}.
\end{equation}

\subsection{Radiative Corrections to $\bar{\theta}$}
\label{sec:radcorr_nbr}

To characterize loop corrections to the renormalizable, tree-level part of the Lagrangian, we use 
\begin{equation}
    \Delta \bar{\theta} = \arg{\det \mathcal{M}_U }.
    \label{eq:theta-barr-full}
\end{equation}
Next, we define $\mathcal{M}_U \equiv \mathcal{M}_0 + \mathcal{M}_1$, where $\mathcal{M}_0$ is the tree-level up quark mass matrix and $\mathcal{M}_1$ characterizes loop corrections to tree-level physics, such that $\mathcal{M}_0^{-1} \mathcal{M}_1 \ll 1$. This lets us write
\begin{equation}
    \begin{aligned}
\operatorname{det} \mathcal{M}_U & =\operatorname{det} \mathcal{M}_0 \operatorname{det}\left(\mathbb{1}+\mathcal{M}_0^{-1} \mathcal{M}_1\right) \\
& \simeq \operatorname{det} \mathcal{M}_0\left(1+\operatorname{tr}\left(\mathcal{M}_0^{-1} \mathcal{M}_1\right)\right). 
\end{aligned}
\end{equation}
Plugging this into Eq.~\ref{eq:theta-barr-full}, we get
\begin{equation}
    \Delta \bar{\theta} \simeq \arg \left(1+\operatorname{tr}\left(\mathcal{M}_0^{-1} \mathcal{M}_1\right)\right) \simeq \operatorname{Im}\left(\operatorname{tr}\left(\mathcal{M}_0^{-1} \mathcal{M}_1\right)\right).
\end{equation}
Finally, if we take the form of $\mathcal{M}_1$ to be 
\begin{equation}
    \mathcal{M}_1 = \left(\begin{array}{cc}
    \mathcal{M}_{\bar{u} u}^{(1)} & \mathcal{M}_{\bar{u} U}^{(1)} \\
    \mathcal{M}_{\bar{U} u}^{(1)} & \mathcal{M}_{\bar{U} U}^{(1)}
    \end{array}\right),
\end{equation}
then we can write $\Delta \bar{\theta}$ as
\begin{equation}
    \Delta \bar{\theta} \simeq \operatorname{Im}\left(m_u^{-1} \mathcal{M}_{\bar{u} u}^{(1)}+\frac{1}{\mu_U}\left(\mathcal{M}_{\bar{U} U}^{(1)} - F m_u^{-1} \mathcal{M}_{\bar{u} U}^{(1)}\right)\right), \label{eq: Theta_Mass_contribution}
\end{equation}

\begin{figure}[t]
\centering
\includegraphics[width=0.5\textwidth]{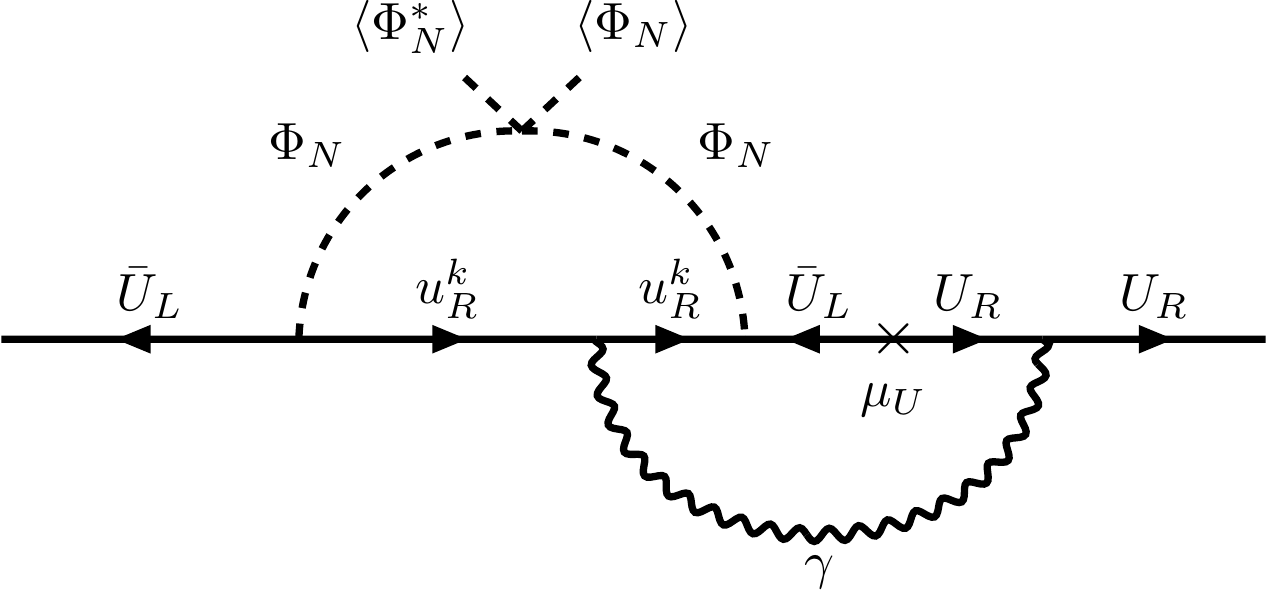}
\caption{A $\bar{U}_L \rightarrow U_R$ two-loop correction that contributes to $\bar{\theta}$.}
\label{fig:UbarU_2loop_diag}
\end{figure}

As discussed in \cite{PhysRevD.99.035014}, in a minimal NB sector like in the BBP model, the possible one-loop contributions to $\bar{\theta}$ are all zero. This is essentially because of the two BSM Yukawas present in the NB Lagrangian. The first loop corrections that contribute to $\bar{\theta}$ correspond to two-loop diagrams like that found in Fig.~\ref{fig:UbarU_2loop_diag}. Computing the diagram where the heavy VLQs couple to the photon, we find that
\begin{equation}
    \mathcal{M}_{\bar{U} U}^{(1)} \supset -i \frac{\mu_U}{ (4 \pi)^3} \qty(\frac{2}{3})^2 2^4 \alpha_{\textrm{EM}} g_{\textrm{clock}}^2 y_k^U \tilde{y}_k^U \cos{2 \theta_N},
\end{equation}
where $g_{\textrm{clock}}^2 < 1$ is the clockwork scalar 4-point vertex coupling and $\alpha_{\textrm{EM}}$ is the electromagnetic coupling "constant". The corresponding contribution to $\bar{\theta}$ is then
\begin{equation}
    \Delta \bar{\theta} \sim \frac{1}{9 \pi^3} \alpha_{\textrm{EM}} g_{\textrm{clock}}^2 y_k^U \tilde{y}_k^U \cos{2 \theta_N}.
\end{equation}
Assuming $\alpha_{\textrm{EM}} \sim \mathcal{O}(10^{-2})$, $g_{\textrm{clock}}^2 \sim \mathcal{O}(10^{-1})$, and $\cos{2 \theta_N} \sim \mathcal{O}(1)$, we find that the constraint $\Delta \bar{\theta} \lesssim 10^{-10}$ leads us to the result:
\begin{equation}
    |y^U| \sim |\tilde{y}^U| \lesssim 10^{-2}.
\end{equation}
\section{Quality Problems}
\label{sec:quality}

We now address several problems associated with the theory presented in Sec.~\ref{sec:nbrelaxion}. Hidden in $\mathcal{L}_{\textrm{NBR}}$ are higher-dimensional operators which, when left unattended, perturb the value of $\Delta \bar{\theta}$ beyond the point of suitability. These operators lead to a quality problem in both the NB and clockwork sectors. In the NB sector, this was first highlighted in \cite{Choi_1993} and later explored further in \cite{dine_challenges_2015}. The quality problem was discussed for clockwork sectors with and without $\mathbb{Z}_2$ symmetries in \cite{higaki_quality_2016}, and was also discussed in the work that introduced the NBR \cite{PhysRevD.99.035014}. Let us examine both quality problems, starting with the clockwork sector.

\subsection{Clockwork Sector}
\label{sec:cw_sector_qp}
 
There is well-known lore that says quantum gravity in general does not respect any global symmetry. This suggests that, for the clockwork chain, Planck-suppressed higher-dimensional operators can spoil the desired symmetry-breaking pattern unless they are sufficiently controlled. These exact difficulties are highlighted in \cite{Higaki_2016, PhysRevD.99.035014}. The operators that must be controlled follow the generalized form $\frac{\Phi_0^{d}}{M_{\textrm{Pl}}^{d - 4}}$ and generate terms like

\begin{equation}
  \Delta V_{\textrm{grav}} \sim \frac{f^d}{M_{\textrm{Pl}}^{d - 4}} \cos(\frac{\phi}{f}),
\end{equation} 

\noindent in the relaxion sector of the theory, where $d$ is an integer representing the mass dimension of the operators. We require that the magnitude of these additional terms be less than or equal to the magnitude of the Higgs-dependent contributions $V_{\textrm{br}}$ found in Eq.~\ref{eq:vbr}, and in doing so, we force the inequality

\begin{equation} \label{eq:quality_cond}
  \frac{f^d}{M_{\textrm{Pl}}^{d - 4}} \cos(\frac{\phi}{f}) \lesssim  \frac{M_{\textrm{br}}^2 v^2}{2} \cos(\frac{\phi}{f}). 
\end{equation}

\noindent Ignoring $\mathcal{O}(1)$ numerical factors, this can be solved for the dimension $d$:

\begin{equation}
   d \gtrsim 4 \frac{\log(\frac{\Lambda_{\textrm{br}}}{M_{\textrm{Pl}}})}{\log(\frac{f}{M_{\textrm{Pl}}})}.
  \label{eq:quality_clockwork_dimension}
\end{equation}

\noindent Furthermore, by using the lower bound $M_{\textrm{br}} \gtrsim v / r_{\textrm{br}}$, the fact that $M_{\textrm{br}} = \Lambda_{\textrm{br}}^2 / v$, the relation $f(\Lambda_H)$ from Eq.~\ref{eq:lambdah}, and the lower bound $\Lambda_H \gtrsim 5$ TeV, we find that $d \gtrsim 6$ which indicates that the NBR theory can only be viable if effective gravitational operators below dimension six are forbidden. A way this may be done is through an ultraviolet (UV) theory where the $\mathrm{U}(1)^{N+1}$ symmetry is an accidental symmetry of the infrared (IR) \cite{PhysRevD.93.085007, coy_dynamical_2017}.

\subsection{Nelson-Barr Sector}
\label{sec:nb_sector_qp}

Since the SM and the additional NB sector are an effective description of a UV theory, higher-dimensional operators allowed by the symmetries of the infrared theory can appear, suppressed by a scale $\Lambda_{\text{EFT}}$. The literature in \cite{dine_challenges_2015, PhysRevD.107.115012} demonstrates that such operators appear in the NB sector. Proceeding from the NB Lagrangian presented in Sec.~\ref{sec:nb_sector}, three dimension-5 operators emerge:

\begin{equation}
\Phi_N \Phi^*_N \bar{U}_LU_R, \quad \Phi_N \bar{Q}_{L,i} \tilde{H} U_R, \quad \text{and} \quad \Phi_N^* \bar{Q}_{L,i} \tilde{H} U_R.
\label{eq:dim-5-quality}
\end{equation}

\noindent Each has an additional contribution to the $\Delta \bar{\theta}$ parameter of Sec.~\ref{sec:nbrelaxion}. The first of these operators perturbs the parameter by an amount

\begin{equation}
\Delta \bar{\theta} \sim \frac{1}{\mu_U \Lambda_{\textrm{EFT}}} \langle \Phi_N^* \Phi_N \rangle,
  \label{eq:nb-quality-dim5-one}
\end{equation}

\noindent where $\Lambda_{\textrm{EFT}}$ provides the mass scale or cutoff for the effective field theory. If one were to assume that $\sqrt{\langle \Phi_N^* \Phi_N \rangle} \sim f \sim \mu_U$ then simple algebra would lead us to $\Delta \bar{\theta} \sim \frac{f}{\Lambda_{\textrm{EFT}}}$. By taking the cutoff to be at the Planck scale ($\Lambda_{\textrm{EFT}} = M_{\textrm{Pl}} \approx 1.22 \times 10^{19}$ GeV) and by enforcing the familiar constraint $\Delta \bar{\theta} \lesssim 10^{-10}$, the scale of $CP$ violation is then restricted to $f \lesssim 10^{9}$ GeV.

\noindent The second and third operators of Eq.~\ref{eq:dim-5-quality} contribute

\begin{equation}
  \Delta \bar{\theta} \sim \Im\qty(\frac{\sqrt{2}   \abs{Y^u}^{-1}}{v \Lambda_{\textrm{EFT}}} \langle \Phi_N \rangle \langle \tilde{H} \rangle) \quad \text{and} \quad \Delta \bar{\theta} \sim \Im\qty(\frac{\sqrt{2} \abs{Y^u}^{-1}}{v \Lambda_{\textrm{EFT}}} \langle \Phi_N^* \rangle \langle \tilde{H} \rangle).
\end{equation}

\noindent This results in $\Delta \bar{\theta} \sim \frac{f}{\Lambda_{\textrm{EFT}}}   \abs{Y^u}^{-1} \sin{\theta_N}$ (the second and third operators result in the same expression modulo a negative sign). Assuming again that $\Lambda_{\textrm{EFT}} = M_{\textrm{Pl}}$, $  \abs{Y^u}^{-1} \sim \mathcal{O}(10)$, and $\sin{\theta_N} \sim \mathcal{O}(1)$, we find that $f \lesssim 10^{8}$ GeV. This is a contentious feature of the NBR. In the next subsection we shall see why.

\subsubsection{Thermal Leptogenesis and Domain Walls}
\label{sec:therm_lepto}

Thermal leptogenesis is a model of baryogenesis which depends on a lepton-anti-lepton asymmetry to generate a baryon-anti-baryon asymmetry \cite{FUKUGITA198645}. Essentially, right-handed neutrinos $N$ are added to the SM and a mass for the SM neutrinos is generated via the type-I seesaw mechanism \cite{MINKOWSKI1977421}. When the right-handed neutrinos decay, they violate $CP$ and lepton number symmetries, giving rise to a net lepton number. The net lepton number is then partially converted to a net baryon number. The resulting baryon asymmetry is directly proportional to
\begin{equation}
  \varepsilon \equiv \frac{\Gamma_{N \rightarrow H \nu} - \Gamma_{N \rightarrow H \bar{\nu}}}{\Gamma_{N \rightarrow H \nu} + \Gamma_{N \rightarrow H \bar{\nu}}},
\end{equation}
where $\Gamma_{N \rightarrow H \nu}$ ($\Gamma_{N \rightarrow H \bar{\nu}}$) is the $N$ decay rate to SM Higgs and neutrinos (anti-neutrinos) \cite{COVI1996169}. In \cite{DAVIDSON200225}, an upper bound was put on $\varepsilon$ by using the most general form of Yukawa matrices which generate the neutrino Dirac mass \cite{CASAS2001171}. We assume a hierarchy in the neutrino masses such that the SM neutrino masses have $m_3 \gg m_1$, and the right-handed neutrinos have $M_3, M_2 \gg M_1$; thus, after solving for the decay rate(s) of right-handed neutrinos to SM Higgs and neutrinos (or anti-neutrinos), the result is that 
\begin{equation}
  \varepsilon \lesssim \frac{3}{8 \pi} \frac{M_1 m_3}{v_h^2},
\end{equation}
where $v_h^2 = v^2 \sin^2{\beta}$ and $v = 246$ GeV. Note that this result is taken from \cite{DAVIDSON200225}, where $\beta$ is defined using $\tan{\beta} = v_2 / v_1$, the ratio of VEVs of the two Higgs doublets associated with the supersymmetric model in which their thermal leptogenesis model is embedded (this is done to stabilize the Higgs mass against quadratic loop divergences). Note that although our model is not supersymmetric, we can still obtain the same results as in the supersymmetric case, but with $v_h \rightarrow v$. An easy way to quell the quadratic loop divergences in the NBR model is to assume that they get absorbed into the Higgs mass scanning parameter $\mu^2(\phi)$. 

For thermal production, the reheating temperature must be high enough to produce the lightest right-handed neutrino, so $T_{\textrm{reh}} \gtrsim M_1$. In the hierarchical-neutrino scenario, the Davidson--Ibarra asymmetry bound then gives the optimistic requirement
\begin{equation} \label{eq:di_bound}
  T_{\textrm{reh}} \gtrsim M_1 \gtrsim 10^8 \,\mathrm{GeV}.
\end{equation}
Detailed estimates of thermal production and washout can strengthen the reheating-temperature requirement to roughly $10^{10}\,\mathrm{GeV}$; the precise value is scenario-dependent.
In the past, $CP$ has been considered as an exact gauge symmetry in the UV \cite{Strominger1985, PhysRevLett.69.2030, Choi_1993}; a consequence of having a spontaneously broken discrete symmetry (like one has in the NB sector) is that the phase transition responsible for such a breaking will cause there to be domain walls separating regions of different vacua \cite{Zeldovich:1974uw, TWBKibble_1976, Zurek1985}. \cite{mcnamara2022reflectionsparitybreaking} concluded that if $CP$ is a gauge symmetry in the UV, spontaneously broken in the IR, then the domain walls which form from the phase transition are exactly stable. This means that in order to dissipate the domain walls, inflation must happen after the spontaneous breaking of $CP$ symmetry. Furthermore, assuming reheating after inflation occurs through perturbative inflaton decay \cite{ABBOTT198229, PhysRevLett.48.1437, DOLGOV1982329}, the authors of \cite{PhysRevD.107.115012} obtain $T_{\textrm{reh}} \lesssim H_{\textrm{inf}} \lesssim f \lesssim 10^8$ GeV, where the last inequality is the NB-quality ceiling. 

If one wants to extend the NBR using a seesaw mechanism to make it mesh with thermal leptogenesis, in order to explain baryogenesis, then there is an alarming issue. The NB quality problem forces the allowable parameter space for the scale of $CP$ violation $f$ to be incompatible with the DI bound. The NBR model as is must depend on some other mechanism to explain baryogenesis. However, it has been shown that by gauging a chiral symmetry, the tension of the NB quality problem with the DI bound can be relaxed \cite{PhysRevD.107.115012}. We now turn to the chiral Nelson--Barr relaxion ($\chi$NBR) model, which gauges a chiral symmetry to relax this tension.
\section{A Chiral Nelson-Barr Relaxion}
\label{sec:chiral_nb_relaxion}

Let us now examine a new model which successfully addresses the difficulties explored in Sec.~\ref{sec:quality}. The model is described by the following Lagrangian 
\begin{equation}
    \mathcal{L}_{\textrm{$\chi$NBR}} = \mathcal{L}_{\textrm{SM}} + \mathcal{L}_{\textrm{R}} + \mathcal{L}_{\textrm{$\chi$NB}} + \mathcal{L}_{\textrm{clock}}.
\end{equation}
Relative to the NBR Lagrangian in Sec.~\ref{sec:nbrelaxion}, we modify the NB sector. We choose to have only one NB Yukawa term, instead of two, with coupling $y_i^U$. The $\chi$NB potential then contains
\begin{equation} \label{eq:snbr}
    V_{\textrm{$\chi$NB}} \supset y^U_i \Phi_N \bar{U}_L u^i_R + (Y^u)^i_{\ j} \bar{Q}_{L,i} \tilde{H} u^j_R + \textrm{H.c.}.
\end{equation}
This sector is the single-scalar version of the NB sectors used in \cite{dine_challenges_2015, PhysRevD.107.115012} and, as we will see in Sec.~\ref{sec:nbyukcons}, causes loop contributions to $\bar{\theta}$ to appear at one-loop order. Taken together with the SM, clockwork, and relaxion sectors, we get a minimal NBR model that would work perfectly were it not for the NB quality problem. 

To alleviate this issue, we gauge a chiral $\mathrm{U}(1)_{\chi}$ symmetry, following the construction in \cite{PhysRevD.107.115012}. In addition to the potential above in Eq.~\ref{eq:snbr}, we introduce a $\mathrm{U}(1)_{\chi}$-charged scalar $\rho$ with two Yukawa couplings:
\begin{equation}
    V_{\textrm{$\chi$NB}} \supset y_U \rho \bar{U}_L U_R + y_B \rho^* \bar{B}_L B_R.
\end{equation}
Note that the VLQ $B$ is added to cancel anomalies and the scalar $\rho$ develops a VEV that generates VLQ masses $\mu_U \equiv y_U \ev{\rho}$ and $\mu_B \equiv y_B \ev{\rho}$. The addition of a mass term for the VLQ $B$ results in an additional diagonal entry in the NB quark mass matrix $\mathcal{M}_0$, but precisely because the additional entry is a diagonal one, there are no additional mass matrix effects to be taken into account. Unlike the NB sector, the renormalizable part of the relaxion and clockwork sectors remains the same. Combining these sectors with the gauged $\mathrm{U}(1)_{\chi}$ symmetry yields a model belonging to the class of $\chi$NBR models.

In the NB sector, there is only one extra Yukawa coupling added to the SM that couples the VLQ $U$ to the NB scalar $\Phi_N$ and right-handed up quarks $u_R$, rather than two, as in the BBP and NBR models. This change ruins cancellations which cause one-loop contributions to $\Delta \bar{\theta}$ to evaluate to zero, as we discuss in Sec.~\ref{sec:nbyukcons}. However, the reduction of Yukawas doesn't modify the expressions for $\Lambda_H$ and $r_{\textrm{roll}}$ very much. Having only one BSM Yukawa amounts to removing the vertex $-i \tilde{y}_i^U \Phi_N^* \bar{U}_L u^i_R$ and its Hermitian conjugate from our theory. Essentially, this causes insertions of $y^U_i \tilde{y}^U_j$ to change to $y^U_i y^U_j$ such that
\begin{equation}\label{eq:newlambdah}
    \Lambda_H \approx \frac{\sqrt{y_i^{U} y_j^{U} \qty(Y^{u \dag} Y^u)_{ij}}}{4 \pi} f,
\end{equation}
and
\begin{equation}\label{eq:newrroll}
    r_{\textrm{roll}} \approx \frac{4 \pi \sqrt{y_k^{U} y_k^{U}}}{y_i^{U} y_j^{U} \qty(Y^{u \dag} Y^u)_{ij}}.
\end{equation}

The $\chi$NBR model also allows for the terms
\begin{equation}
    \mathcal{L}_{\chi\text{NBR}} \supset -(Y^e)^i_j \bar{L}_{L,i} H e_R^j - (Y^{\nu})^i_j \bar{L}_{L,i} \tilde{H} N^j - y^N \Phi_0^* N^i N_i.
\end{equation}
The first term is a Dirac mass term for the charged leptons. The second term generates a mass mixing between SM neutrinos and right-handed neutrinos. Finally, the third operator generates a Majorana mass term for the right-handed neutrinos once the scalar $\Phi_0$ develops a VEV. Together, these three generate a mass matrix for the SM and right-handed neutrinos via the type-I seesaw mechanism \cite{MINKOWSKI1977421}. Assuming that the $N$ particles are in thermal equilibrium with the SM bath and therefore that their Majorana mass is below $T_{\text{reh}}$, their decay (which violates $CP$ symmetry and lepton number) generates a net lepton number, which is then partially converted into a net baryon asymmetry. Thus, the $\chi$NBR addresses the strong $CP$ problem through the NB sector and the electroweak hierarchy problem through the relaxion sector; with suitable right-handed-neutrino masses and reheating, it can also accommodate thermal leptogenesis.

\subsection{One-loop Constraint on Nelson-Barr Yukawa} \label{sec:nbyukcons}

In this theory, one-loop corrections do not simply evaluate to zero as in the NBR model. Instead, we must evaluate the finite contributions that these make to $\Delta \bar{\theta}$. 

\begin{figure}
\centering
\includegraphics[width=0.5\textwidth]{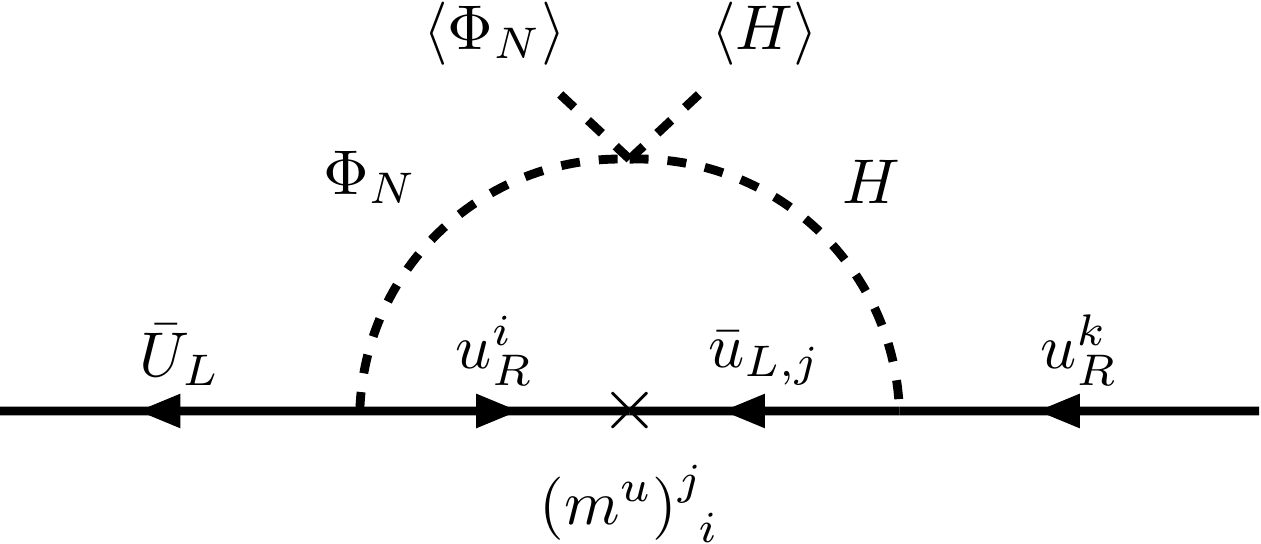}
\caption{$\bar{U}_L \rightarrow u_R$ one-loop correction.}
\label{fig:diag_Ubaru_1loop}
\end{figure}

The relevant one-loop diagram can be found in Fig.~\ref{fig:diag_Ubaru_1loop}. 
The diagram evaluates to 
\begin{equation}
    -i (\mathcal{M}_{\bar{U} u}^{(1)})_i \approx -i \frac{\sqrt{2}}{32 \pi^2 f} y^U_j (m^u)^{k}_{\ i} (m^u)^{k}_{\ j} \log(\frac{f^2}{m_h^2}) \mathrm{e}^{i \theta_N}.
\end{equation}
Its contribution to $\bar{\theta}$ is then
\begin{equation}
    \Delta \bar{\theta} \sim \frac{v}{\mu_U} \frac{\lambda}{32 \sqrt{2} \pi^2} (Y^u)^i_{\ j} y^U_i y^U_j \log(\frac{f^2}{m_h^2}) \sin{2 \theta_N}.
\end{equation}
Assuming that $v = 246$ GeV, $\mu_U \sim 1$ TeV, $\lambda \sim \mathcal{O}(10^{-1})$, $\abs{Y^u} \sim \mathcal{O}(10^{-1})$, $\log(\frac{f^2}{m_h^2}) \sim \mathcal{O}(10)$, and $\sin(2 \theta_N) \sim \mathcal{O}(1)$, we find that the constraint $\Delta \bar{\theta} \lesssim 10^{-10}$ leads to the order-of-magnitude bound $|y^U|\lesssim10^{-3}$. For the phenomenological analysis below, we adopt the more restrictive case $|y^U|\lesssim10^{-4}$, which lies below this bound. The smallness of the Yukawa is technically natural since taking the coupling strength to zero yields a $U(1)_N$ symmetric NB sector. 

To finish up our business with loops, we calculate the one-loop correction to the parameter $\lambda$. Below the scale of $CP$ violation $f$, the scalar field $\Phi_N$ develops a VEV and one-loop corrections to $\lambda$ lead to the roll potential of the relaxion sector (Eq.~\ref{eq:vroll}). However, if one considers energies above the scale $f$, then $\Phi_N$ has a VEV equal to zero and the correction to $\lambda$ is 
\begin{equation}
    \delta \lambda \sim \frac{1}{16 \pi^2} y_i^U y_j^U \qty(Y^{u \dag} Y^u)_{ij} \log\qty(\frac{\Lambda^2_{\textrm{UV}}}{f^2}).
\end{equation}
Note that we use a hard momentum cutoff and integrate from $k=f$ to $k=\Lambda_{\textrm{UV}}$. We would like to keep $\lambda \lesssim 10^{-1}$, so we have that $\qty|\delta \lambda| \lesssim \lambda$ and
\begin{equation}
    10^{-1} \gtrsim \frac{1}{16 \pi^2} y_i^U y_j^U \qty(Y^{u \dag} Y^u)_{ij} \log(\frac{\Lambda^2_{\textrm{UV}}}{f^2}).
\end{equation}
Plugging in $|y^U| \sim \mathcal{O}(10^{-4})$, $\abs{Y^u} \sim \mathcal{O}(10^{-1})$, and assuming $\log(\frac{\Lambda^2_{\textrm{UV}}}{f^2}) \sim \mathcal{O}(10^2)$, we get that
\begin{equation}
    10^{-1} \gtrsim \frac{1}{16 \pi^2} y_i^U y_j^U \qty(Y^{u \dag} Y^u)_{ij} \log(\frac{\Lambda^2_{\textrm{UV}}}{f^2}) \sim 6 \times 10^{-11},
\end{equation}
which is manifestly true. Thus, the couplings in the theory are consistent with $\bar{\theta} \lesssim 10^{-10}$.

\subsection{Anomaly Cancellation} 
\label{sec:anom_canc}

Because the introduction of the $\mathrm{U}(1)_{\chi}$ symmetry causes the VLQ $U$ to behave chirally, there exist uncanceled anomalies that must be addressed. One way to ensure that these additional anomalies vanish is to add extra fermions charged under the $\mathrm{U}(1)_{\chi}$ symmetry. This introduces contributions to the preexisting anomalies that leads to their cancellation. \\
Our anomaly-cancellation construction follows \cite{PhysRevD.107.115012}. We use a NB sector which is essentially the Hermitian conjugate of theirs. We also utilize a two-parameter solution where the parameters are $x$ and $y$ and correspond to the $\mathrm{U}(1)_{\chi}$ charges of the Higgs doublet $H$ and the scalar $\rho$. The $\mathrm{U}(1)_{\chi}$ charge of the $\Phi_0$ field $Q_0$ is fixed to be 2. 

After gauging the $\mathrm{U}(1)_{\chi}$ symmetry, a difficulty in assigning charges to the clockwork scalar fields arises due to the fact that we need terms in the clockwork sector that go as $\Phi_j^{\dag} \Phi_{j+1}^q$ to break the $\mathrm{U}(1)^{N+1}$ symmetry down to the desired $\mathrm{U}(1)_N$ symmetry. If we naïvely made all $\mathrm{U}(1)_{\chi}$ charges for the $\Phi_j$ scalars the same, then we would mistakenly forbid the needed symmetry breaking terms which make the clockwork mechanism work. Instead, to allow the desired terms, we require that the $\mathrm{U}(1)_{\chi}$ charges $Q_j$ of the scalars be such that $Q_j = q Q_{j+1}$. This means that the charge of the $N$-th site is $Q_N = Q_0 q^{-N}$. Large $N$ can lead to extremely small charges at the $N$-th site of the clockwork chain. Depending on how strongly one considers the consequences of the weak gravity conjecture \cite{Arkani-Hamed_2007} this could be a problem, but we don't consider it to be fatal for our construction. 

Another possibility for assigning charges could be to make the $\mathrm{U}(1)_{\chi}$ charge of $\Phi_N$ to be $Q_N = 2$, then the charge of $\Phi_0$ would be $Q_0 = Q_N q^N$. We would also have to allow for the operator $y^N \Phi_N^* N^i N_i$, instead of $y^N \Phi_0^* N^i N_i$. However, this poses a problem that is extremely jarring, as the charges of scalars lower down the clockwork chain can become dangerously large depending on the number of sites $N$. In general, large charges (or gauge couplings) lead to faster running of gauge couplings, which can lead to Landau poles being at much lower scales than desired. 

In making our charge assignments, we enforce the existence of the operator $(Y^{\nu})^i_j \bar{L}_{L,i} \tilde{H} N^j$, which in turn forces the $\mathrm{U}(1)_{\chi}$ charges of the right-handed neutrinos to be equal to 1. Because of this, the scalar responsible for generating the right-handed neutrino Majorana mass through its VEV must carry a charge of magnitude 2. 

A convenience of the charge assignment we chose is that it forbids the existence of operators of the form $\Phi_0^d$. This solves the clockwork quality problem, but $\bar{\theta}$ contributing dimension-5 operators must be forbidden if we are also to solve the NB quality problem discussed in Sec.~\ref{sec:nb_sector_qp}. We find that we must enforce $y \neq 0$ and $y \neq 4q^{-N}$ to forbid them. In Table~\ref{tab:CNBRcharges}, we detail the charge assignments of each field in our theory for $x = 0$, $y = 1$, which is one of the simplest solutions. The table lists the charge assignments for the fields shown; the intermediate clockwork-site charges follow $Q_j=qQ_{j+1}$. Note that we take $q = 3$ in this work.
\begin{table}[h!]
\centering
\caption{Charge assignments for the fields under 
$\mathrm{SU}(3)_c \times \mathrm{SU}(2)_L \times \mathrm{U}(1)_Y \times \mathrm{U}(1)_{\chi}$.}
\label{tab:CNBRcharges}
\begin{tabular}{lcccc}
\hline\hline
& $\mathrm{SU}(3)_c$ & $\mathrm{SU}(2)_L$ & $\mathrm{U}(1)_Y$ & $\mathrm{U}(1)_{\chi}$ \\
\hline
$\bar{Q}_{L, i}$ & $\mathbf{\bar{3}}$ & $\mathbf{2}$ & $-1/6$ & $+1/3$ \\
$u_R^i$ & $\mathbf{3}$ & $\mathbf{1}$ & $+2/3$ & $-1/3$ \\
$d_R^i$ & $\mathbf{3}$ & $\mathbf{1}$ & $-1/3$ & $-1/3$ \\
$H$ & $\mathbf{1}$ & $\mathbf{2}$ & $+1/2$ & $0$ \\
$\bar{U}_L$ & $\mathbf{\bar{3}}$ & $\mathbf{1}$ & $-2/3$ & $+1/3-2q^{-N}$ \\
$U_R$  & $\mathbf{3}$ & $\mathbf{1}$ & $+2/3$ & $-4/3+2q^{-N}$ \\
$\bar{B}_L$ & $\mathbf{\bar{3}}$ & $\mathbf{1}$ & $+2/3$ & $-1/3+2q^{-N}$ \\
$B_R$ & $\mathbf{3}$ & $\mathbf{1}$ & $-2/3$ & $+4/3-2q^{-N}$ \\
$\rho$ & $\mathbf{1}$ & $\mathbf{1}$ & $0$ & $+1$ \\
$\Phi_0$ & $\mathbf{1}$ & $\mathbf{1}$ & $0$ & $+2$ \\
$\Phi_N$ & $\mathbf{1}$ & $\mathbf{1}$ & $0$ & $+2q^{-N}$ \\
$e_{R,i}$ & $\mathbf{1}$ & $\mathbf{1}$ & $-1$ & $+1$ \\
$\bar{L}_{L,i}$ & $\mathbf{1}$ & $\mathbf{2}$ & $+1/2$ & $-1$ \\
$N_i$ & $\mathbf{1}$ & $\mathbf{1}$ & $0$ & $+1$ \\
\hline\hline
\end{tabular}
\end{table}

\subsection{On the Clockwork and Nelson-Barr Quality Problems}
\label{sec:on_cw_nb_qp}

We now explain how the charge assignment addresses the clockwork and NB quality problems discussed in Secs.~\ref{sec:cw_sector_qp} and~\ref{sec:nb_sector_qp}. By assigning chiral charges across the model, we forbid both the clockwork sector dimension-$d$ operators of the form $\Phi_0^d$ and the NB sector dimension-5 operators that give rise to the naïve bound $f \lesssim 10^8$ GeV. Thus, the leading clockwork-breaking and dimension-five NB operators are forbidden by this charge assignment. In the NB sector, there will still be operators of mass dimension 6 and above which can contribute to $\bar{\theta}$. However, as we will see, this does not pose a problem. The additional $\mathrm{U}(1)_{\chi}$ symmetry permits a few dimension-6 operators which give the leading contributions to $\bar{\theta}$:
\begin{equation}
    \mathcal{S}_2 \rho \bar{U}_L U_R \quad \text{and} \quad \mathcal{S}_2 \bar{Q}_{L,i} \tilde{H} u_R^j, 
\end{equation}
where $\mathcal{S}_2$ is any gauge-invariant scalar combination of mass dimension 2. For example, $\mathcal{S}_2 = \rho^* \rho, \Phi_0^* \Phi_0, \Phi_N^* \Phi_N$. Out of the above operators, it is the first type which has the largest possible contribution to $\bar{\theta}$. We can see this through the contribution of the operator $\qty(\Phi_N^* \Phi_N) \rho \bar{U}_L U_R$:
\begin{equation}
    \Delta \bar{\theta} \simeq \Im\qty[ \frac{1}{\mu_U \Lambda^2_{\textrm{EFT}}} \ev{\Phi^*_N \Phi_N} \ev{\rho} ] \sim \frac{1}{y_U} \frac{f^2}{\Lambda^2_{\textrm{EFT}}},
\end{equation}
where we assumed $\mu_U \sim f$ and $\ev{\Phi^* \Phi} \sim f^2$. If we further assume that $y_U \sim \mathcal{O}(1)$ and $\Lambda_{\textrm{EFT}} \sim M_{\textrm{Pl}}$, then the condition that $\Delta \bar{\theta} \lesssim 10^{-10}$ yields $f \lesssim 10^{13}$ GeV. The rest of the operators yield results that are either the same or at greater values. Note that $H^{\dag} H \rho \bar{U}_{L} U_R$ is also a possible operator, but because $\ev{H} \sim v$ it will be subleading. This relaxes the tension with thermal leptogenesis. Combining the reheating-temperature requirement in Eq.~\ref{eq:di_bound} with the residual NB-quality ceiling gives a potentially viable interval, subject to the reheating and inflation assumptions stated above.
\section{Phenomenology}
\label{sec:pheno}

As discussed in Sec.~\ref{sec:chiral_nb_relaxion}, the chiral symmetry raises the NB-quality ceiling from $f \lesssim 10^{8}\,\mathrm{GeV}$ to $f \lesssim 10^{13}\,\mathrm{GeV}$. Under the assumptions stated there, the one-loop estimate gives $|y^U| \lesssim 10^{-3}$. We compare the phenomenology at $|y^U| \sim 10^{-3}$ and $|y^U| \sim 10^{-4}$. As we will see, the viable parameter space for $|y^U|\sim10^{-3}$ is smaller than that for $|y^U|\sim10^{-4}$, motivating our use of the latter value for $|y^U|$. We now derive these constraints on the relaxion parameter space.

\noindent Suppose that the lower bound on $\Lambda_H$, the scale of new physics, is $\Lambda_H \gtrsim$ 5 TeV. By assuming an MFV scenario and using the relation $\sum_{ij} \left[ Y^{u \dag} Y^u \right]_{i j} \approx \sum_{ij} y_t^{2} V^*_{3 i} V_{3 j} \approx 2 (\frac{m_t}{v})^{2} \cdot 10^{-2} \sim \mathcal{O}(10^{-2})$ \cite{isidori_minimal_2012}, where $m_t = $ 174 GeV and $v = $ 246 GeV, we find that
\begin{equation}
    f_{\textrm{min}} \approx 6 \times 10^9 \text{ GeV} \cdot \qty(\frac{10^{-4}}{\abs{y^{U}}}) \qty(\frac{\Lambda^{\textrm{min}}_H}{\text{5 TeV}}).
    \label{eq:fmin}
\end{equation}

\subsection{Relaxion-Higgs Mixing Angle}

Through couplings present in the relaxion sector, a relaxion-Higgs mixing angle is generated. This quantity can be found from the eigenvalues of the relaxion-Higgs mass matrix, which, in terms of the $\chi$NBR potential $V_{\chi \text{NBR}}$, is 
\begin{equation}
    \mathcal{M}_{h \phi} = \pmqty{M_{h h} & M_{h \phi} \\ M_{\phi h} & M_{\phi \phi}} = \eval{\pmqty{\partial_{h h} V_{\chi \text{NBR}} & \partial_{h \phi} V_{\chi \text{NBR}} \\ \partial_{\phi h} V_{\chi \text{NBR}} & \partial_{\phi \phi} V_{\chi \text{NBR}}}}_{h \rightarrow v, \phi \rightarrow \phi_0 }
    \label{eq:relaxion-higgs-matrix}
\end{equation} 
\noindent By requiring that the physical Higgs mass $m_h$ correspond to one of the eigenvalues such that $\det(\mathcal{M}_{h \phi} - m_h^2 \mathbb{1}) = 0$, we find that
\begin{equation}
    \mathcal{M}_{h \phi} = \pmqty{\frac{M_{h \phi}^2}{M_{\phi \phi} - m_h^2} + m_h^2 & M_{h \phi} \\ M_{\phi h} & M_{\phi \phi}},
\end{equation}
such that $M_{hh} = \frac{M_{h \phi}^2}{M_{\phi \phi} - m_h^2} + m_h^2$.

The corresponding eigenvalues are 
\begin{equation}
    \lambda_h = m_h^2 \qquad \textrm{and} \qquad \lambda_{\phi} = \frac{M_{h \phi}^2}{M_{\phi \phi} - m_h^2} + M_{\phi \phi} = m_{\phi}^2.
\end{equation}

\noindent Now, using the relaxion-Higgs matrix elements, calculated explicitly in Appendix~\ref{sec:calculation_relaxion-higgs_matrix}, our final expression for the relaxion mass becomes 

\begin{equation}
    m_{\phi}^2 \approx \frac{\Lambda_{\textrm{br}}^4}{2 f^2} \qty( \cos(\frac{\phi_0}{f}) - \frac{2 \Lambda_{\textrm{br}}^4}{v^2 m_h^2} \sin^2\qty(\frac{\phi_0}{f})) \approx \frac{\Lambda_{\textrm{br}}^4}{2 f^2}.
    \label{eq:relaxion-mass}
\end{equation}

\noindent By leveraging Eq.~\ref{eq:relaxion-mass} and following the analysis provided in Appendix~\ref{sec:calculation_sin}, the mixing angle is found to be 

\begin{equation}
        \sin{\theta}
        \approx 2 \times 10^{-15} \left( \frac{10^{-4}}{\abs{y^{U}}} \right) \left( \frac{\Lambda_H}{\text{5 TeV}} \right) \left( \frac{m_{\phi}}{\mathrm{eV}} \right)^2.
        \label{eq:relaxion-higgs-mixing-pheno}
\end{equation}

\begin{figure}
\centering
\begin{subfigure}{0.49\textwidth}
    \centering
    \includegraphics[width=\textwidth]{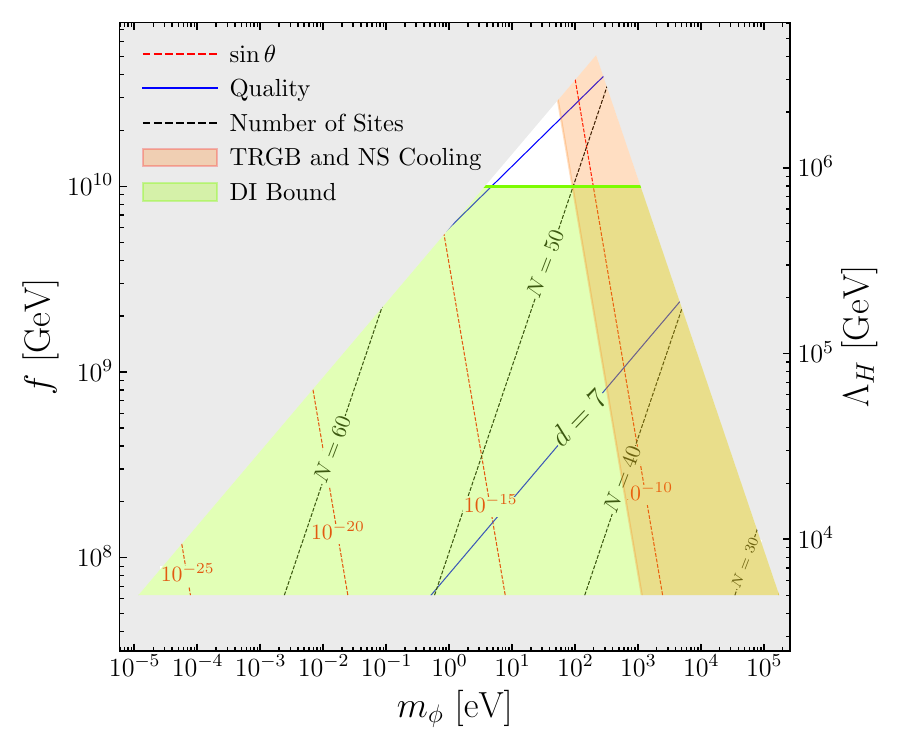}
    \caption{$\qty|y^U| \sim \qty|\tilde{y}^U| \sim 10^{-2}$}
    \label{fig:ypsi1e-02_smyuk1e-01_mass_plot}
\end{subfigure}
\begin{subfigure}{0.49\textwidth}
    \centering
    \includegraphics[width=\textwidth]{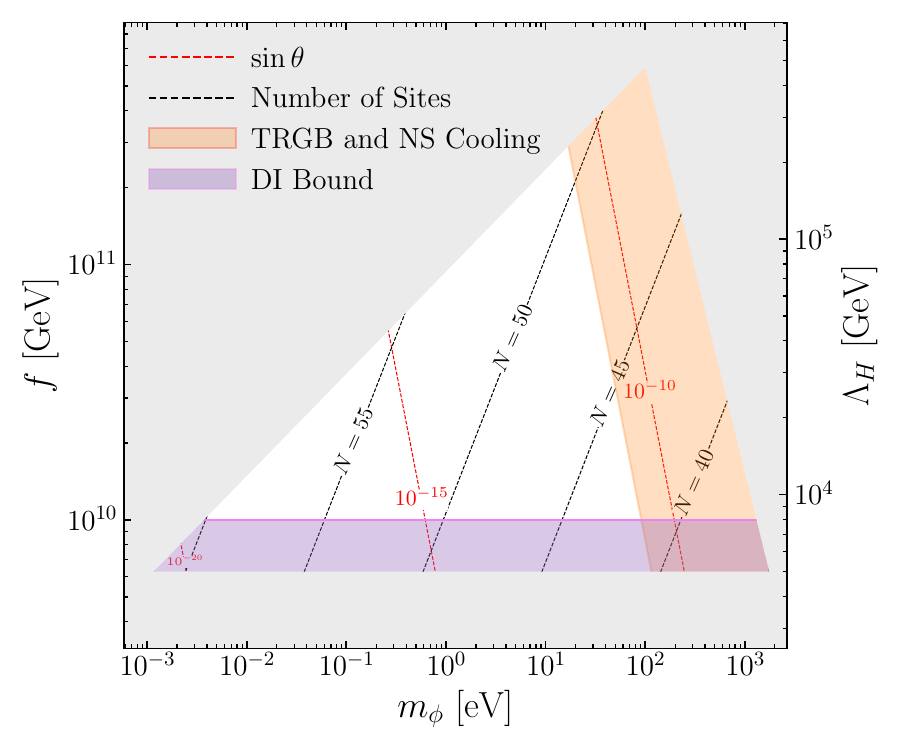}
    \caption{$\qty|y^U| \sim 10^{-4}$}
    \label{fig:ypsi1e-04_smyuk1e-01_mass_plot}
\end{subfigure}
\caption{Available parameter space plots. In Fig.~\ref{fig:ypsi1e-02_smyuk1e-01_mass_plot} is the parameter space for the NBR model and in Fig.~\ref{fig:ypsi1e-04_smyuk1e-01_mass_plot} is the parameter space for the $\chi \textrm{NBR}$ model. The red dashed curves show the relaxion-Higgs mixing angle as a function of the relaxion mass and the black dashed curves show the number of required clockwork sites as a function of the relaxion mass. The relaxion-Higgs mixing angle constraint (labeled "TRGB and NS Cooling") comes from combining tip of the red giant branch and neutron star cooling bounds \cite{tlqz-713s, debru2026newboundslightscalars}, while the "DI Bound" is the reheating temperature requirement for thermal leptogenesis ($T_{\textrm{reh}} \gtrsim 10^8-10^{10}\,\mathrm{GeV}$, with the stronger bound plotted) \cite{DAVIDSON200225}. The blue curves in the NBR model parameter space plot indicate the dimension of effective gravitational operators which satisfy the inequality in Eq. ~\ref{eq:quality_cond} for a given relaxion mass $m_{\phi}$ and scale of $CP$ violation $f$.}
\label{fig:ypsi1e-04_mass_plot}
\end{figure}

\noindent Curves of constant $\sin{\theta}$ are plotted in red on Fig.~\ref{fig:ypsi1e-02_smyuk1e-01_mass_plot} and Fig.~\ref{fig:ypsi1e-04_smyuk1e-01_mass_plot}.
 
\subsection{Relaxion Mass bounds}

An upper bound for the relaxion mass $m_\phi$ can be found by utilizing Eq.~\ref{eq:relaxion-mass}. Using the relation $\Lambda_{\textrm{br}} \approx \sqrt{M_{\textrm{br}}v} $, coupled with the upper bound on the back reaction scale $M_{\textrm{br}} \lesssim m^2_h / v$, as well as $f_{\textrm{min}}$ of Eq.~\ref{eq:fmin}, we find

\begin{equation}
    m_{\phi}^{\textrm{max}} \approx 2 \textrm{ keV} \cdot \left( \frac{\abs{y^{U}}}{10^{-4}} \right) \left( \frac{5 \textrm{ TeV}}{\Lambda_H^{\textrm{min}}} \right).
\end{equation}

\noindent From conditions on the Hubble scale of inflation and the stopping of the slow-rolling of the relaxion  \cite{PhysRevLett.115.221801, PhysRevD.99.035014} we get the upper bound

\begin{equation}
    \Lambda_H \lesssim \left( \frac{M_{\textrm{Pl}}}{r_{\textrm{roll}}} \right)^{1/2} \left( \frac{\Lambda_{\textrm{br}}^4}{f} \right)^{1/6}.
\end{equation}

\noindent  With this bound, along with a few lines of algebra, we arrive at the minimum for the relaxion mass:

\begin{equation}
    m_{\phi}^{\textrm{min}} \approx 10^{-3} \textrm{ eV} \cdot \left( \frac{10^{-4}}{\abs{y^{U}}} \right) \left( \frac{\Lambda_H^{\textrm{min}}}{5 \textrm{ TeV}} \right)^{5/2}.
\end{equation}

\begin{figure}
\centering
\centering
\includegraphics[width=0.6\textwidth]{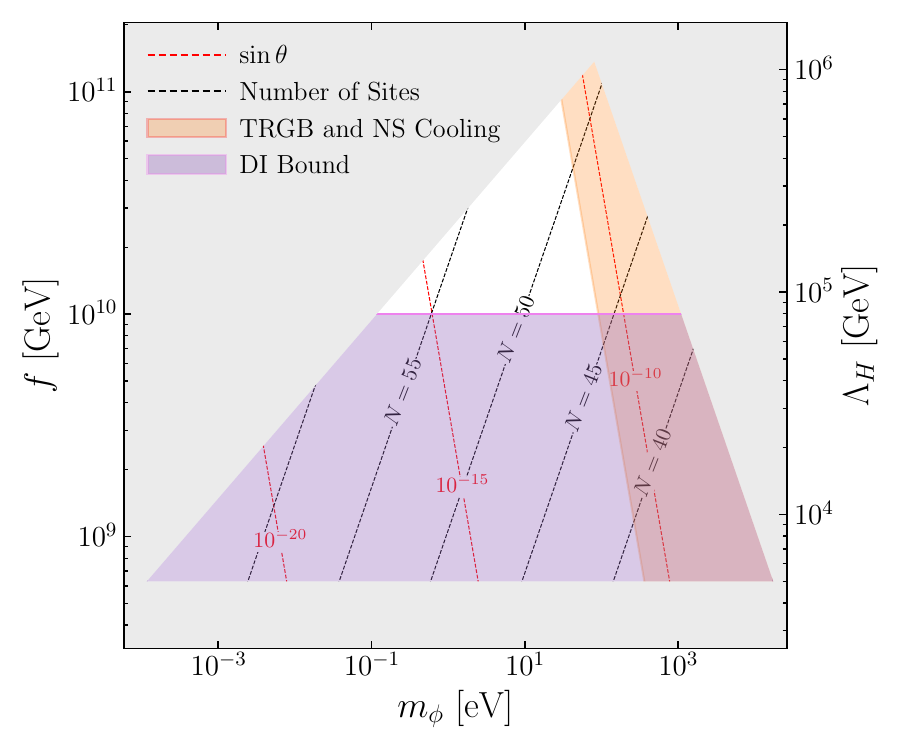}
\caption{Available $\chi$NBR parameter space for $\abs{y^U} \sim 10^{-3}$. The constraints and curves plotted are the same as those in Fig. ~\ref{fig:ypsi1e-04_smyuk1e-01_mass_plot}.}
\label{fig:ypsi1e-03_smyuk1e-01_mass_plot}
\end{figure}

The intersection point where $m_{\phi}^{\textrm{min}} = m_{\phi}^{\textrm{max}}$ defines the phenomenological ceiling $f_{\textrm{max}}^{\rm pheno}$ for the scale of $CP$ violation $f$. Letting $m_{\phi}^{\textrm{min}}$ and $m_{\phi}^{\textrm{max}}$ be functions of $\Lambda_H$, we can set them equal and solve for $f$, yielding
\begin{equation}
    f_{\textrm{max}}^{\rm pheno} \sim 4 \times 10^{11} \text{ GeV} \left( \frac{10^{-4}}{\abs{y^{U}}} \right)^{3/7}.
\end{equation}
If we were working within the framework of the NBR, we would have that $\abs{y^{U}} \lesssim 10^{-2}$. This results in a more constrained bound:
\begin{equation}
    f_{\textrm{max}}^{\rm pheno} \sim 5 \times 10^{10} \text{ GeV} \left( \frac{10^{-2}}{\abs{y^{U}}} \right)^{3/7}.
\end{equation}
By removing a BSM Yukawa from the NB sector of the NBR model, we make one-loop contributions non-vanishing, which drives down the allowed values for $\abs{y^{U}}$ and yields a phenomenological ceiling $f_{\textrm{max}}^{\rm pheno}$ about eight times that in the NBR case. A comparison of the different bounds on $f$ between the NBR and $\chi$NBR models is shown in Table~\ref{tab:f_bounds}. While the increase in $f$ is substantial, the allowed mass range of the relaxion is also increased to span six orders of magnitude, as opposed to roughly three (accounting for the DI bound). As seen in Fig.~\ref{fig:ypsi1e-03_smyuk1e-01_mass_plot}, the allowed parameter space for $\abs{y^U} \sim 10^{-3}$ seems more restrictive than that for $\abs{y^U} \sim 10^{-4}$, which is exactly why we choose $\abs{y^U} \sim 10^{-4}$.

\begin{table}[t]
    \centering
    \caption{Upper limits on the scale of $CP$ violation $f$ in the NBR and $\chi$NBR models. The overall limit is set by the more restrictive of the quality and phenomenological limits. The DI bound is shown for comparison; it is a lower bound on the reheating temperature, not an upper bound on $f$.}
    \label{tab:f_bounds}
    \resizebox{.95\textwidth}{!}{%
    \begin{tabular}{lccc}
        \toprule
        Model
        & NB-quality ceiling
        & Phenomenological ceiling
        & Overall limit \\
        \midrule
        NBR
        & $f \lesssim 10^8\,\mathrm{GeV}$
        & $f \lesssim 5 \times 10^{10}\,\mathrm{GeV}$
        & $T_{\mathrm{reh}} \lesssim f \lesssim 10^8\,\mathrm{GeV}$ \\
        $\chi$NBR
        & $f \lesssim 10^{13}\,\mathrm{GeV}$
        & $f \lesssim 4 \times 10^{11}\,\mathrm{GeV}$
        & $T_{\mathrm{reh}} \lesssim f \lesssim 4 \times 10^{11}\,\mathrm{GeV}$ \\
        DI bound
        & ---
        & ---
        & $T_{\mathrm{reh}} \gtrsim 10^{8\text{--}10}\,\mathrm{GeV}$ \\
        \bottomrule
    \end{tabular}
        }
\end{table}

\subsection{Quality Mass relation}

By using $\sqrt{2} f m_{\phi} \approx \Lambda_{\textrm{br}}^2$ in conjunction with Eq.~\ref{eq:quality_clockwork_dimension}, we find 

\begin{equation}
         f^{d - 2} \gtrsim 2 m^2_{\phi} M_{\textrm{Pl}}^{d - 4}.
\end{equation}

\noindent We can use $f = 4 \pi \Lambda_H /\sqrt{y_i^{U} y_j^{U} \qty(Y^{u \dag} Y^u)_{ij}}$ to put the Higgs effective cutoff $\Lambda_H$ in terms of the dimension $d$ of the effective gravitational operators. We obtain

\begin{equation}
    \Lambda_H \gtrsim 10^{13} \textrm{ GeV} \cdot \qty(\frac{\abs{y^{U}}}{10^{-4}}) \qty(\frac{2 m^2_{\phi}}{M^2_{\textrm{Pl}}})^{\frac{1}{d-2}}. \label{eq:lambda_quality_not_simplified}
\end{equation}

\noindent Curves of constant $d$ are plotted in blue in Fig.~\ref{fig:ypsi1e-02_smyuk1e-01_mass_plot}. The chiral symmetry of the $\chi$NBR model removes the clockwork-quality constraint discussed here; the blue curves therefore only apply to the original NBR model.

\subsection{Clockwork Site Number Mass Dependence}

The relaxion comes to a stop at the  EW VEV \cite{PhysRevD.99.035014}, which means that by minimizing the potential with respect to the Higgs and relaxion fields, $h$ and $\phi$, respectively, we obtain

\begin{equation}
    \frac{\Lambda_H}{\Lambda_{\textrm{br}}} \approx \left ( -\frac{F}{2 r^2_{\textrm{roll}} f} \right )^{1/4}.
\end{equation}

\noindent Given $\Lambda_{\textrm{br}}^2 = \sqrt{2} m_{\phi} f$, the  result $\frac{F}{f} = 3^N$ \cite{PhysRevD.93.085007}, Eq.~\ref{eq:newrroll}, and Eq.~\ref{eq:fmin}, we solve for $\Lambda_H$ and get that

\begin{equation} \label{eq:lambdahnumsites}
    \Lambda_H \approx 10^{-8} \textrm{ GeV} \cdot 3^{N/2} \cdot \left ( \frac{m_{\phi}}{\textrm{eV}} \right ).
\end{equation}

\noindent Curves of constant $N$ are present in Figs.~\ref{fig:ypsi1e-02_smyuk1e-01_mass_plot} \&~\ref{fig:ypsi1e-04_smyuk1e-01_mass_plot} and are given by the black dashed lines.
\section{Conclusion}
\label{sec:conc}

The NBR is a model composed of three new sectors in addition to the SM: a relaxion sector $\mathcal{L}_{\text{R}}$, a clockwork sector $\mathcal{L}_{\text{clock}}$, and a NB sector $\mathcal{L}_{\text{NB}}$. In the relaxion sector, the relaxion dynamically scans the Higgs mass, eventually coming to rest at its observed value; coupled with a model of slow-roll inflation, this solves the EW hierarchy problem. However, the relaxion sector alone gives no explanation for how the hierarchy of scales $F \gg f$, which it requires to function, is realized. This task is left to the clockwork sector, which, through a chain of $N+1$ scalars with nearest-neighbor couplings and the corresponding breaking of $\mathrm{U}(1)^{N+1}$ down to $\mathrm{U}(1)_N$, generates the necessary separation of scales. Together, the clockwork and relaxion sectors can compose a successful relaxion model, and if one further identifies the relaxion with the azimuthal mode of the scalar field introduced in minimal NB models, a NBR is realized, which additionally solves the strong $CP$ problem. 

NBR models, however, face quality problems in both the clockwork and NB sectors. Following the techniques of \cite{PhysRevD.107.115012}, we gauged a chiral $\mathrm{U}(1)_\chi$ symmetry and introduced additional VLQs to cancel the anomalies that arise from making the VLQ $U$ chiral, which in turn forbids the dangerous operators responsible for both the clockwork and NB quality problems. We also added right-handed neutrinos, whose couplings generate a mass for the SM neutrinos via the type-I seesaw mechanism. With suitable right-handed-neutrino masses and reheating, the model can accommodate thermal leptogenesis as a mechanism for generating the baryon asymmetry. Altogether, our model solves the EW hierarchy problem, the strong $CP$ problem, and the problem of baryogenesis, while improving on the original NBR by raising the allowed scale of $CP$ violation to a level compatible with the reheating-temperature requirement inferred from the Davidson--Ibarra bound, within the assumptions studied here.

Looking forward, it would be interesting to explore whether the supersymmetric UV-completion techniques of \cite{dine_challenges_2015, Choi_2016} could be extended to construct a full UV completion of our $\chi$NBR or a similar model. It could also be useful to further explore the consequences of the weak gravity conjecture on our construction, as we only mentioned it briefly in Sec.~\ref{sec:anom_canc}. We ponder what kinds of predictions our model's right-handed neutrino sector could make for the $CP$-violating phase of the Pontecorvo-Maki-Nakagawa-Sakata (PMNS) matrix. Testing any predictions, however, awaits the resolution of the current tension between the T2K and NO$\nu$A measurements of the PMNS $CP$-violating phase \cite{PhysRevD.103.112008, PhysRevD.106.032004}. Finally, a useful direction for future work is a detailed analysis of experimental signatures for our model and similar models. 

\appendix

\section{Diagonalization of Mass Matrix}
\label{sec:pdiag_mm}

\subsection{Method 1}

This first method of diagonalization is taken from \cite{cherchiglia_flavor_2021} \& \cite{cherchiglia_consequences_2020}. The NB Lagrangian looks like
\begin{equation}
	V_{\textrm{NB}} \supset \mu_U \bar{U}_L U_R + \qty( y^U_i \Phi_N + \tilde{y}^U_i \Phi_N^{*} ) \bar{U}_L u^i_R + (Y^u)^i_{\ j} \bar{Q}_{L,i} \tilde{H} u^j_R + \textrm{H.c.}.
\end{equation}
Letting $F_i(\Phi_N) = y_i^U \Phi_N + \tilde{y}_i^U \Phi_N^{*}$ (or, one could make the choice: $F_i(\Phi_N) = y_i^U \Phi_N$), it can be rewritten as
\begin{equation}
	V_{\textrm{NB}} \supset \mu_U \bar{U}_L U_R + F_i(\Phi) \bar{U}_L u^i_R + (Y^u)^i_{\ j} \bar{Q}_{L,i} \tilde{H} u^j_R + \textrm{H.c.}.
\end{equation}
A generic Lagrangian involving VLQs is 
\begin{equation}
	V_{\textrm{NB}}^{\textrm{gen}} \supset M \bar{U}_L U_R + (Y^U)^i \bar{Q}_{L,i} \tilde{H} U_R + (y^u)^{i}_{\ j} \bar{Q}_{L,i} \tilde{H} u^j_R + \textrm{H.c.}.
\end{equation}
After the scalar field $\Phi_N$ acquires a VEV, the NB Lagrangian takes the form
	\begin{equation}
		V_{\textrm{NB}} \supset \pmqty{ \bar{u} & \bar{U} }_L \underbrace{ \pmqty{ \frac{v}{\sqrt{2}} Y^u & 0 \\ F & \mu_U } }_{ \mathcal{ M }_0 } \pmqty{ u \\ U }_R,
	\end{equation}
while the generic Lagrangian takes the following form
	\begin{equation}
		V_{\textrm{NB}}^{\textrm{gen}} \supset \pmqty{ \bar{u} & \bar{U} }_L \underbrace{ \pmqty{ \frac{v}{\sqrt{2}} y^u & \frac{v}{\sqrt{2}} Y^U \\ 0 &  M } }_{ \mathcal{ M }'_0 } \pmqty{ u \\ U }_R.
	\end{equation}
We seek a matrix $W_R$ which rotates the first form of the mass matrix $\mathcal{ M }_0$ into the second $\mathcal{ M }_0'$ such that
	\begin{equation}
		\mathcal{ M }_0 W_R = \mathcal{ M }'_0
	\end{equation}
We can write the matrix $W_R$ as four blocks such that 
	\begin{equation}
		W_R = \pmqty{ K & R \\ S & T }.
	\end{equation}
Now, we have that 
	\begin{equation}
		\pmqty{ \frac{v}{\sqrt{2}} Y^u & 0 \\ F & \mu_U } \pmqty{ K & R \\ S & T } = \pmqty{ \frac{v}{\sqrt{2}} y^u & \frac{v}{\sqrt{2}} Y^U \\ 0 &  M } \label{eq:main}.
	\end{equation}
Doing some finagling, we find four equations
	\begin{align}
		y^u & = Y^u K, \label{eq:upyuk} \\
		  Y^U & = Y^u R, \\
		S & = -\frac{ F K }{ \mu_U }, \\
		M & = F R + \mu_U T \label{eq:meq}.		
	\end{align}
From the unitarity of $W_R$, from which follows the relation, $\mathcal{ M }_0 \mathcal{ M }_0^{ \dag } = \mathcal{ M } \mathcal{ M }^{ \dag }$, we obtain the following sum rule:
	\begin{equation}
		y^u (y^u)^{ \dag } + Y^U (Y^U)^{ \dag } = Y^u (Y^u)^{ \dag }.
	\end{equation}
Unitarity of $W_R$ also means that 	
	\begin{equation}
		M^2 = H_F \equiv F F^{ \dag } + \mu_U^2
	\end{equation}
and 
	\begin{equation}
		K K^{ \dag } = \mathbb{1} - \frac{ F^{ \dag } F }{ M^2 }. \label{eq:kkdag}
	\end{equation}
Using Eq.~\ref{eq:meq} and its square, we see that the simplest consistent choices for $R$ and $T$ are
	\begin{equation}
		R = \frac{ F^{ \dag } }{ M } \quad \text{and} \quad T = \frac{ \mu_U }{ M }.
	\end{equation}
Finally, using Eqs.~\ref{eq:upyuk} \&~\ref{eq:kkdag}, we obtain 
	\begin{align*}
		y^u y^{ u \dag } & = Y^u K K^{ \dag } Y^{ u \dag }, \\
		& = Y^u \qty(\mathbb{1} - \frac{ F^{ \dag } F }{M^2}) (Y^u)^{ \dag },
	\end{align*}
	\begin{equation}
		\frac{ v^2 }{ 2 } y^u y^{ u \dag } =  m_u m_u^{ \intercal } - \frac{ m_u F^{ \dag } F m_u^{ \intercal } }{ F F^{ \dag } + \mu_U^2 }
	\end{equation}

\subsection{Method 2}

This second method is taken from \cite{valenti_ckm_2021}. The NB Lagrangian looks like
	\begin{equation}
		V_{\textrm{NB}} \supset \mu_U \bar{U}_L U_R + F_i(\Phi_N) \bar{U}_L u^i_R + (Y^u)^i_{\ j} \bar{Q}_i \tilde{H} \bar{u}^j_R + \textrm{H.c.}.
	\end{equation}
It takes the following matrix form:
	\begin{equation}
		V_{\textrm{NB}} \supset \pmqty{ \bar{u} & \bar{U} }_L \pmqty{ m_u & 0 \\ F & \mu_U } \pmqty{ u \\ U }_R,
	\end{equation}
with $m_u = \frac{v}{\sqrt{2}} Y^u$. Before $\Phi_N$ acquires a VEV, this mass matrix, call it $\mathcal{M}_0$, has determinant zero, which means that the CP violating phase $\bar{ \theta } = 0$ at tree level. Now, we know that for a matrix with distinct left and right eigenvectors, it will be diagonalized by the bi-unitary transformation  $V_L^{ \dag } \mathcal{M}_0 V_R = \text{diag}(\bar{m}, \bar{M})$, where $\bar{m}$ and $\bar{M}$ are the SM quark masses and the heavy quark mass in the diagonal basis. We also know that $V_L^{ \dag } \mathcal{M}_0 \mathcal{M}_0^{ \dag } V_L = \text{diag}(\bar{m}^2, \bar{M}^2)$. This means that 
	\begin{align*}
		V_L^{ \dag } \mathcal{M}_0 \mathcal{M}_0^{ \dag } V_L  & = \text{diag}(\bar{m}^2, \bar{M}^2), \\
		\mathcal{M}_0 \mathcal{M}_0^{ \dag } V_L  & = V_L \text{diag}(\bar{m}^2, \bar{M}^2).
	\end{align*}
We can conveniently write the matrix $V_L$ in block form as
	\begin{equation}
		V_L = \pmqty{ K & R \\ S & T }.
	\end{equation}
This lets us write 
	\begin{align*}
		\mathcal{M}_0 \mathcal{M}_0^{ \dag } V_L  & = V_L \text{diag}(\bar{m}^2, \bar{M}^2) \\
		\pmqty{ m_u & 0 \\ F & \mu_U } \pmqty{ m_u^{\intercal} & F^{ \dag } \\ 0 & \mu_U^* } \pmqty{ K & R \\ S & T } & = \pmqty{ K & R \\ S & T } \pmqty{ \bar{m}^2 & 0 \\ 0 & \bar{M}^2 }, \\
		\pmqty{ m_u m_u^{\intercal} & m_u F^{ \dag } \\ F m_u^{\intercal} & F F^{ \dag } + \mu_U^2 } \pmqty{ K & R \\ S & T } & = \pmqty{ K \bar{m}^2 & R \bar{M}^2 \\ S \bar{m}^2 & T \bar{M}^2 }, \\
		\pmqty{ m_u m_u^{\intercal} K + m_u F^{ \dag } S & m_u m_u^{\intercal} R + m_u F^{ \dag } T \\ F m_u^{ \intercal } K + (F F^{ \dag } + \mu_U^2) S & F m_u^{ \intercal } R + (F F^{ \dag } + \mu_U^2) T } & = \pmqty{ K \bar{m}^2 & R \bar{M}^2 \\ S \bar{m}^2 & T \bar{M}^2 }.
	\end{align*}
This leads to four equations:
	\begin{align}
		m_u m_u^{\intercal} K + m_u F^{ \dag } S & = K \bar{m}^2, \label{eq:ksk} \\
		m_u m_u^{\intercal} R + m_u F^{ \dag } T & = R \bar{M}^2, \label{eq:rtr} \\
		F m_u^{\intercal} K + (F F^{ \dag } + \mu_U^2) S & = S \bar{m}^2, \label{eq:kss} \\
		F m_u^{\intercal} R + (F F^{ \dag } + \mu_U^2) T & = T \bar{M}^2. \label{eq:rtt}
	\end{align}
Taking Eq.~\ref{eq:kss}, we get
	\begin{equation*}
		S = -F m_u^{\intercal} K \qty(F F^{ \dag } + \mu_U^2 - \bar{m}^2)^{-1}.
	\end{equation*}
Using $F F^{ \dag } + \mu_U^2 = M^2$ and using the approximation $\bar{m}^2 / M^2 << 1$, we get that 
	\begin{equation}
		S \approx -\frac{ F m_u^{ \intercal } K }{ M^2 } \qty( 1 + \frac{ \bar{m}^2 }{ M^2 } ). \label{eq:s}
	\end{equation}
By inserting Eq.~\ref{eq:s} into Eq.~\ref{eq:ksk}, we find that 
	\begin{align*}
		K \bar{m}^2 & = m_u m_u^{\intercal} K - \frac{ m_u F^{ \dag } F m_u^{ \intercal } K }{ M^2 } \qty( 1 + \frac{ \bar{m}^2 }{ M^2 } ), \\
		K \bar{m}^2 K^{ -1 } & = m_u m_u^{\intercal} - \frac{ m_u F^{ \dag } F m_u^{ \intercal } }{ M^2 } \qty( 1 + \frac{ \bar{m}^2 }{ M^2 } ), 
	\end{align*}
	\begin{equation}
		K \bar{m}^2 K^{ -1 } \approx m_u m_u^{\intercal} - \frac{ m_u F^{ \dag } F m_u^{ \intercal } }{ F F^{ \dag } + \mu_U^2 }. 
	\end{equation}
Note that $K$ is the usual CKM matrix and is given as 
	\begin{equation}
		K = \pmqty{ 1 & 0 & 0 \\ 0 & c_{23} & s_{23} \\ 0 & -s_{23} & c_{23} } \pmqty{ c_{13} & 0 & s_{13} e^{-i \delta} \\ 0 & 1 & 0 \\ -s_{13} e^{i \delta} & 0 & c_{13} } \pmqty{ c_{12} & s_{12} & 0 \\ -s_{12} & c_{12} & 0 \\ 0 & 0 & 1 },
	\end{equation}
where $c_{ i j } = \cos{ \theta_{ i j } }$ and $s_{ i j } = \sin{ \theta_{ i j } }$. This yields the same result as the first method (as it should), since $K \bar{m}^2 K^{ -1 } = \frac{ v^2 }{ 2 } y^u y^{ u \dag }$.
\section{Details of the Clockwork Mechanism}
\label{sec:det_clockmech}

The clockwork model of \cite{PhysRevD.93.085007} has $N+1$ complex scalars $\Phi_j$, where $j = 0, \dots, N$. Their potential is described by 
\begin{equation} \label{eq:vclock_app}
    V_{\textrm{clock}}(\Phi) = \sum_{j = 0}^N \qty( -m^2 |\Phi_j|^2 + g^2_{\textrm{clock}} |\Phi_j|^4 ) + \Delta V_{\textrm{clock}}(\Phi),
\end{equation}
where 
\begin{equation}
    \Delta V_{\textrm{clock}}(\Phi) = \sum_{j = 0}^{N - 1} \qty( \varepsilon \Phi_j^{\dag} \Phi_{j + 1}^3 + \textrm{H.c.} ).
\end{equation}

The term in parentheses in Eq.~\ref{eq:vclock_app} respects a global $\mathrm{U}(1)^{N+1} = \mathrm{U}(1)_0 \times \mathrm{U}(1)_1 \times \dots \times \mathrm{U}(1)_N$ symmetry while $\Delta V_{\textrm{clock}}(\Phi)$ explicitly breaks this symmetry down to the $\mathrm{U}(1)_N$ symmetry. Under the $\mathrm{U}(1)_N$ symmetry, the fields $\Phi_j$ have charges $Q = 1, \frac{1}{3}, \frac{1}{9}, \dots, \frac{1}{3^N}$. If we enforce $\varepsilon \ll g^2_{\textrm{clock}} < 1$, then we can take the limit $\varepsilon \rightarrow 0$ and expand around the scalar VEVs $\abs{\langle \Phi_j \rangle} = \frac{f}{\sqrt{2}} \ \forall \ j$. Below the scale $\sqrt{2} m$ we can assume the fields are all at their VEVs such that
\begin{equation}
    \Phi_j \rightarrow U_j \equiv \frac{f}{\sqrt{2}} \mathrm{e}^{i \pi_j / f}.
\end{equation}
If we keep the terms in $\varepsilon$, then the Lagrangian for the psuedo-Nambu-Goldstone bosons is 
\begin{align*}
    \mathcal{L}_{\textrm{pNGB}} & \approx f^2 \sum_{j = 0}^{N} \partial_\mu U^{\dag}_j \partial^{\mu} U_j + \qty( \varepsilon \sum_{j = 0}^{N - 1} U_j^{\dag} U_{j + 1}^3 + \textrm{H.c.} ), \\
    & \approx \frac{1}{2} \sum_{j = 0}^{N} \partial_{\mu} \pi_j \partial^{\mu} \pi_j + \varepsilon \frac{f^4}{4} \sum_{j = 0}^{N - 1} \mathrm{e}^{i \qty(3 \pi_{j + 1} - \pi_j ) / f} + \textrm{H.c.}.
\end{align*}
If we expand this and take the terms quadratic in the fields we get 
\begin{equation}
    V_{\textrm{clock}}^{(2)} = \frac{1}{8} \varepsilon f^2 \sum_{j = 0}^N \qty( 3 \pi_{j + 1} - \pi_j )^2.
\end{equation}
This results in the following mass matrix:
\begin{equation}
    M_{i j}^2= \frac{1}{4} \epsilon f^2\left(\begin{array}{cccccccc}
1 & -3 & 0 & 0 & & & & \\
-3 & 1+3^2 & -3 & 0 & & \cdot & \cdot & \cdot \\
0 & -3 & 1+3^2 & -3 & & & & \\
0 & 0 & -3 & 1+3^2 & & & & \\
& & & & \cdot & & & \\
\cdot & & & & \cdot & & & \\
\cdot & & & & & \cdot & & \\
\cdot & & & & & 1+3^2 & -3 \\
& & & & & & -3 & 3^2
\end{array}\right)
\end{equation}

In the case of finite $N$, the mass eigenmodes ${a^j}$ satisfying 
\begin{equation}
    (M^2)^{ j l} a_l = \lambda a_j
\end{equation}
consist of a zero mode
\begin{equation}
    a_j^{(0)} = \frac{\mathcal{N}}{3^j},
\end{equation}
where 
\begin{equation}
    j = 0, \dots, N \quad \textrm{and} \quad \mathcal{N} = \frac{1}{\sqrt{\sum_{j = 0}^N \frac{1}{3^{2 j}}}}.
\end{equation}
There are also massive modes of the form
\begin{equation}
    a_j(\theta) = \sin{j \theta} + A \cos{j \theta},
\end{equation}
with eigenvalues
\begin{equation} 
    \lambda(\theta) = 1 + 3^2 -2 \cdot 3 \cos{\theta}.
\end{equation}
The above forms of $a_j(\theta)$ and $\lambda(\theta)$ work for $1 \leq j \leq N - 1$. At the boundaries $j = 0$ and $j = N$, one instead gets
\begin{align*}
    A & = - \frac{\sin{\theta}}{3 - \cos{\theta}} \\
    0 & = \qty(1 + 3^2) \sin(N + 1)\theta - 3 \sin{N \theta} - 3 \sin(N+2)\theta.
\end{align*}
The zero-mode eigenvector is 
\begin{equation}
    (\vec{a}^{(0)})^{\intercal} = \mathcal{N} \qty( 1 \quad \frac{1}{3} \quad \frac{1}{9} \quad \cdots \quad \frac{1}{3^N}  ).
\end{equation}
The zero-mode has an exponentially suppressed overlap $\sim 3^{-N}$ with whatever operator $\Phi_N$ couples to in the UV. 

\section{Nelson-Barr Loop Generation of Roll Sector}
\label{sec:nb_to_rel}

The roll potential is generated at one-loop level from the $\phi\rightarrow\phi$ loop diagram and the $\phi\phi\rightarrow h h$ box diagram. Both of these diagrams are found in Fig.~\ref{fig:feynmandiagrams}. To start off, we first work with the NB sector from the NBR model:
\begin{equation}
    V_{\mathrm{NB}} \supset \left[y_i^U \Phi_N+\tilde{y}_i^U \Phi_N^*\right] \bar{U}_L u_R^i + (Y^u)^i_{\ j} \bar{Q}_{L,i} \tilde{H} u_R^j + \text{H.c.}.
\end{equation}
Substituting in $\Phi_N = \frac{f}{\sqrt{2}} e^{i \frac{\phi}{F}}$ and $\tilde{H}^{\intercal} = \frac{1}{\sqrt{2}} \pmqty{v & 0}$, we get 
\begin{align*}
    V_{\mathrm{NB}} &\supset \left[y_i^U \Phi_N + \tilde{y}_i^U \Phi_N^* \right] \bar{U}_L u_R^i + (Y^u)^i_{\ j} \bar{Q}_{L,i} \tilde{H} u_R^j + \text{H.c.}, \\
    & \supset \left[y_i^U \qty(\frac{f}{\sqrt{2}} e^{i \frac{\phi}{F}}) + \tilde{y}_i^U \qty(\frac{f}{\sqrt{2}} e^{-i \frac{\phi}{F}}) \right] \bar{U}_L u_R^i + (Y^u)^i_{\ j} \bar{Q}_{L,i} \tilde{H} u_R^j + \text{H.c.}, \\
    & \supset \frac{f}{\sqrt{2}} \left[ y_i^{U} e^{i \frac{\phi}{F}} + \tilde{y}_i^{U} e^{-i \frac{\phi}{F}} \right] \bar{U}_L u_R^i + (Y^u)^i_{\ j} \bar{Q}_{L,i} \tilde{H} u_R^j + \text{H.c.}.
\end{align*}
Now, we can expand the exponentials and only keep the terms linear in $\phi$. We get  
\begin{equation}
    V_{\mathrm{NB}} \supset i \frac{f}{\sqrt{2} F} \left[ y_i^{U} - \tilde{y}_i^{U} \right] \phi \bar{U}_L u_R^i + (Y^u)^i_{\ j} \bar{Q}_{L,i} \tilde{H} u_R^j + \text{H.c.}.
\end{equation}
If we had used the NB sector from the $\chi \textrm{NBR}$ model, which is the same as in the NBR model except for an additional term proportional to $\tilde{y}_i^{U}$, we would've gotten
\begin{equation} \label{CNBR}
    V_{\mathrm{NB}} \supset i \frac{f}{\sqrt{2} F} y_i^{U}\phi \bar{U}_L u_R^i + (Y^u)^i_{\ j} \bar{Q}_{L,i} \tilde{H} u_R^j + \text{H.c.}.
\end{equation}
Using the results for the NB sector in the NBR model (if one wants to consider the $\chi \textrm{NBR}$ model NB sector, one just needs to use Eq.~\ref{CNBR}), we can read off the factors that correspond to vertices in the loop diagram found in Fig.~\ref{fig:diag_phiphi}. We will discuss how to find the vertices of the loop diagram in Fig.~\ref{fig:diag_phiphihh} afterwards. Note that we will use the same conventions as in \cite{Dreiner_2010}. The loop amplitude $i I_1$ computed in the low external momentum limit with hard-momentum cutoff at loop momentum $k=f$ is 
\begin{align*}
    -i I_1 & = -\qty(-i \frac{f}{\sqrt{2} F} y_k^{U}) \qty(-i \frac{f}{\sqrt{2} F} \tilde{y}_k^{U}) \int \frac{\mathrm{d}^4 k}{(2 \pi)^4} \Tr[\frac{i k \cdot \bar{\sigma}}{\qty(k^2 - m^2_{u, i})} \frac{i (p - k) \cdot \sigma}{\qty((p - k)^2 - \mu_U^2)}], \\
    & = -\frac{f^2}{2 F^2} y_k^{U} \tilde{y}_k^{U} \int \frac{\mathrm{d}^4 k}{(2 \pi)^4} \frac{\Tr[\bar{\sigma}^{\mu} \sigma^{\nu}] k_{\mu} k_{\nu}}{\qty(k^2 - m^2_{u, i}) \qty(k^2 - \mu_U^2)}, \\
    & \approx -\frac{f^2}{F^2} y_k^{U} \tilde{y}_k^{U} \int \frac{\mathrm{d}^4 k}{(2 \pi)^4} \int_0^1 \mathrm{d} x \frac{k^2}{\qty[x \qty(k^2 - m^2_{u, i}) + (1-x) \qty(k^2 - \mu_U^2)]^2}, \\
    & \approx -\frac{f^2}{F^2} y_k^{U} \tilde{y}_k^{U} \int \frac{\mathrm{d}^4 k}{(2 \pi)^4} \int_0^1 \mathrm{d} x \frac{k^2}{\qty[k^2 - M^2]^2}, \\
    & \approx -i \frac{f^4}{16 \pi^2 F^2} y_k^{U} \tilde{y}_k^{U}.
\end{align*}
Now, since the loop-corrected mass is $m^2 = m_0^2 + I_1$, where $m_0$ is the bare mass, we must get an additional term in the potential that looks like $\frac{1}{2} I_1 |\phi|^2$
\begin{align*}
    V_{\textrm{roll}} \supset & -\Lambda_{\textrm{roll}}^4 \cos(\frac{\phi}{F}) = -r_{\textrm{roll}}^2 \Lambda_H^4 \cos(\frac{\phi}{F}), \\
    \approx & -r_{\textrm{roll}}^2 \Lambda_H^4 \qty(1 - \frac{1}{2} \frac{|\phi|^2}{F^2}) \approx \frac{1}{2} \frac{r_{\textrm{roll}}^2}{F^2} \Lambda_H^4 |\phi|^2,
\end{align*}
we must require that
\begin{align*}
    \frac{1}{2} \frac{r_{\textrm{roll}}^2}{F^2} \Lambda_H^4 |\phi|^2 & = \frac{1}{2} I_1 |\phi|^2, \\
    \frac{r_{\textrm{roll}}^2}{F^2} \Lambda_H^4 & = I_1, \\
    r_{\textrm{roll}}^2\Lambda_H^4 & = \frac{f^4}{16 \pi^2} y_k^{U} \tilde{y}_k^{U}, \\
    r_{\textrm{roll}}^2 & = \frac{f^4}{16 \pi^2 \Lambda_H^4} y_k^{U} \tilde{y}_k^{U},
\end{align*}
\begin{equation}\label{eq:rrollmu}
    r_{\textrm{roll}} = \frac{\sqrt{y_k^{U} \tilde{y}_k^{U}}}{4 \pi} \frac{f^2}{\Lambda_H^2}.
\end{equation}
To finish this calculation, we need an expression for $\Lambda_H^2$. For this we must compute the amplitude of the quark loop diagram for the process $\phi \phi \rightarrow h h$. This requires simplifying the SM up type quark-Higgs Yukawa operator $(Y^u)^i_{\ j} Q_{L,i} \tilde{H} \bar{u}_R^j$. Doing so gets us
\begin{equation}
    (Y^u)^i_{\ j} Q_i \tilde{H} \bar{u}_R^j \rightarrow \frac{1}{\sqrt{2}} (Y^u)^i_{\ j} h u_i \bar{u}_R^j.
\end{equation}
The loop, call it $I_2$, is then,
\begin{align*}
    i I_2 & = -\frac{1}{3} \qty(\frac{f}{\sqrt{2} F} y_i^{U}) \qty( \frac{f}{\sqrt{2} F} \tilde{y}_j^{U}) \qty(-i \frac{1}{\sqrt{2}} (Y^{\dag}_u)^k_{\ i}) \qty(-i \frac{1}{\sqrt{2}} (Y^u)^k_{\ j}) \\
    & \quad \times \int \frac{\mathrm{d}^4 k}{(2 \pi)^4} \Tr[\frac{i \bar{\sigma} \cdot k}{k^2 - m^2_{u, i}} \frac{i \sigma \cdot (k-p_3)}{(k-p_3)^2 - m^2_{u, j}} \frac{i \bar{\sigma} \cdot (k-p)}{(k-p)^2 - m^2_{u, k}} \frac{i \sigma \cdot (k-p_1)}{(k-p_1)^2 - \mu^2_U}],
\end{align*}
where $p = p_1 + p_2 = p_3 + p_4$. Note that we've also suppressed additions of $i \varepsilon$ in the denominators of propagators. Working through the trace, we get 
\begin{align*}
    i I_2 & = \frac{f^2}{12 F^2} y_i^{U} \tilde{y}_j^{U} \qty(Y^{u \dag} Y^u)_{ij} \\
    & \quad \times \int \frac{\mathrm{d}^4 k}{(2 \pi)^4} \Tr[\frac{i \bar{\sigma} \cdot k}{k^2 - m^2_{u, i}} \frac{i \sigma \cdot (k-p_3)}{(k-p_3)^2 - m^2_{u, j}} \frac{i \bar{\sigma} \cdot (k-p)}{(k-p)^2 - m^2_{u, k}} \frac{i \sigma \cdot (k-p_1)}{(k-p_1)^2 - \mu^2_U}], \\
    & = \frac{f^2}{12 F^2} y_i^{U} \tilde{y}_j^{U} \qty(Y^{u \dag} Y^u)_{ij} \\
    & \quad \times \int \frac{\mathrm{d}^4 k}{(2 \pi)^4} \frac{\Tr[\bar{\sigma}^{\mu} \sigma^{\nu} \bar{\sigma}^{\sigma} \sigma^{\rho}] k_{\mu} (k - p_3)_{\nu} (k - p)_{\sigma} (k-p_1)_{\rho}}{(k^2 - m^2_{u, i}) ((k-p_3)^2 - m^2_{u, j}) ((k-p)^2 - m^2_{u, k}) ((k-p_1)^2 - \mu^2_U)}, \\
    & = \frac{f^2}{6 F^2} y_i^{U} \tilde{y}_j^{U} \qty(Y^{u \dag} Y^u)_{ij} \\
    & \quad \times \int \frac{\mathrm{d}^4 k}{(2 \pi)^4} \frac{(g^{\mu \nu} g^{\sigma \rho} - g^{\mu \sigma} g^{\nu \rho} + g^{\mu \rho} g^{\nu \sigma} - i \epsilon^{\mu \nu \sigma \rho}) k_{\mu} (k - p_3)_{\nu} (k - p)_{\sigma} (k-p_1)_{\rho}}{(k^2 - m^2_{u, i}) ((k-p_3)^2 - m^2_{u, j}) ((k-p)^2 - m^2_{u, k}) ((k-p_1)^2 - \mu^2_U)}.
\end{align*}
Applying the small external momentum limit, we get that 
\begin{align*}
    i I_2 & \approx -\frac{f^2}{6 F^2} y_i^{U} \tilde{y}_j^{U} \qty(Y^{u \dag} Y^u)_{ij} \\
    & \quad \times \int \frac{\mathrm{d}^4 k}{(2 \pi)^4} \frac{k^4}{(k^2 - m^2_{u, i}) (k^2 - m^2_{u, j}) (k^2 - m^2_{u, k}) (k^2 - \mu^2_U)}.
\end{align*}
Using the Feynman parameter trick for combining four denominators and then evaluating the resulting integral, we see that 
\begin{equation}
    i I_2 \approx -\frac{f^2}{F^2} y_i^{U} \tilde{y}_j^{U} \qty(Y^{u \dag} Y^u)_{ij} \int \frac{\mathrm{d}^4 k}{(2 \pi)^4} \int_0^1 \mathrm{d} x_1 \mathrm{d} x_2 \mathrm{d} x_3 \frac{k^4}{\qty[k^2 - M^2]^4},
\end{equation}
where $M^2 = x_1 m_{u,i}^2 + x_2 m_{u,j}^2 + x_3 m_{u,k}^2 + (1-x_1-x_2-x_3) \mu_U^2$. Doing the rest of this integral yields
\begin{align*}
    i I_2 & \approx -i \frac{f^2}{F^2} y_i^{U} \tilde{y}_j^{U} \qty(Y^{u \dag} Y^u)_{ij} \frac{2 \pi^2}{(2 \pi)^4} \frac{1}{12} \int_0^1 \mathrm{d} x_1 \mathrm{d} x_2 \mathrm{d} x_3 \qty[6 \log(\frac{f^2}{M^2}) - 11], \\
    & \approx -i \frac{f^2}{F^2} y_i^{U} \tilde{y}_j^{U} \qty(Y^{u \dag} Y^u)_{ij} \frac{2 \pi^2}{(2 \pi)^4} \frac{3}{12} \qty[\log(\frac{f^2}{\mu_U^2}) + \frac{1}{3} \qty(2 \pi i- 1)],
\end{align*}
\begin{equation}\label{eq:loop2}
    i I_2 \approx -i \frac{1}{32 \pi^2} \frac{f^2}{F^2} y_i^{U} \tilde{y}_j^{U} \qty(Y^{u \dag} Y^u)_{ij}.
\end{equation}
Note that we approximated $M^2$ by taking $\mu_U \gg m_{u,i}, m_{u,j}, m_{u,k}$ and then integrated with respect to the Feynman parameters. We also take $\mu_U \sim f$, such that the log is $\mathcal{O}(1)$. Now, since a correction to the potential equal to $-I_2 \qty|\phi|^2 h^2$ is produced, we find that
\begin{align*}
    -\Lambda_H^2 \cos(\frac{\phi}{F}) H^\dagger H \rightarrow \frac{\Lambda_H^2}{2 F^2} \qty|\phi|^2 h^2 & = -I_2 \qty|\phi|^2 h^2, \\
    \frac{\Lambda_H^2}{2 F^2} & = -I_2, \\
    \frac{\Lambda_H^2}{2 F^2} & = \frac{1}{32 \pi^2} \frac{f^2}{F^2} y_i^{U} \tilde{y}_j^{U} \qty(Y^{u \dag} Y^u)_{ij}, \\
    \Lambda_H^2 & = \frac{1}{16 \pi^2} y_i^{U} \tilde{y}_j^{U} \qty(Y^{u \dag} Y^u)_{ij} f^2.
\end{align*}
So, 
\begin{equation}
    \Lambda_H \approx \frac{\sqrt{y_i^{U} \tilde{y}_j^{U} \qty(Y^{u \dag} Y^u)_{ij}}}{4 \pi} f.
\end{equation}
Plugging this back into Eq.~\ref{eq:rrollmu}, we find that
\begin{equation}
    r_{\textrm{roll}} \approx \frac{4 \pi \sqrt{y_k^{U} \tilde{y}_k^{U}}}{y_i^{U} \tilde{y}_j^{U} \qty(Y^{u \dag} Y^u)_{ij}}.
\end{equation}
Note that if we had used the $\chi \textrm{NBR}$ model NB sector, these quantities would have only $y_i^U$ and not $\tilde{y}_i^U$, as in Eq.s~\ref{eq:newlambdah}-~\ref{eq:newrroll}.
\section{Relaxion-Higgs Mixing}
\label{sec:relaxion_higgs_mixing}

Through the same routes of analysis taken to investigate the relaxion-Higgs mixing angle, an expression for the relaxion mass can be derived, as shown in Eq.~\ref{eq:relaxion-mass}.  This analysis is done by calculating the matrix elements in the relaxion-Higgs mass matrix $\mathcal{M}_{h\phi}$. This is done below.

\subsection{Calculation of Matrix Elements}
\label{sec:calculation_relaxion-higgs_matrix}

It is the objective of this section to explicitly calculate the elements of the relaxion-Higgs mass matrix in~\ref{eq:relaxion-higgs-matrix}. Each matrix element is expressed as derivatives on the potential $V_{\textrm{NB}}$:

\begin{equation}
    \mathcal{M}_{h \phi} = \pmqty{M_{h h} & M_{h \phi} \\ M_{\phi h} & M_{\phi \phi}} = \eval{\pmqty{\partial_{h h} V_{\chi \text{NBR}} & \partial_{h \phi} V_{\chi \text{NBR}} \\ \partial_{\phi h} V_{\chi \text{NBR}} & \partial_{\phi \phi} V_{\chi \text{NBR}}}}_{h \rightarrow v, \phi \rightarrow \phi_0 }.
    \label{eq:app_relaxion-higgs-matrix}
\end{equation} 

\noindent Before we calculate these elements let us quickly turn to  $\eval{\partial_h V_{\chi \text{NBR}}}_{h \rightarrow v, \phi \rightarrow \phi_0 }$ and $\eval{\partial_{\phi} V_{\chi \text{NBR}}}_{h \rightarrow v, \phi \rightarrow \phi_0 }$ which will become helpful later. Evaluated at the minimum of the potential $V_{\chi \text{NBR}}$, we find that 
\begin{equation*} 
     \eval{\partial_h V_{\chi \text{NBR}}}_{h \rightarrow v, \phi \rightarrow \phi_0 } = \kappa v \Lambda_H^2 - v \Lambda_H^2 \cos(\frac{\phi_0}{F}) + \lambda v^3 - v M_{\textrm{br}}^2 \cos(\frac{\phi_0}{f})  = 0,
\end{equation*}
\begin{equation} 
    \kappa \Lambda_H^2 - \Lambda_H^2 \cos(\frac{\phi_0}{F}) + \lambda v^2 - M_{\textrm{br}}^2 \cos(\frac{\phi_0}{f}) = 0.
    \label{eq:relaxion_mixing_dvh}
\end{equation}
Likewise, 
\begin{equation*} 
     \eval{\partial_{\phi} V_{\chi \text{NBR}}}_{h \rightarrow v, \phi \rightarrow \phi_0} = \left ( \frac{v^2}{2}\Lambda_H^2 + r_\textrm{roll}^2 \Lambda_H^4 \right )\frac{1}{F} \sin(\frac{\phi_0}{F}) + \left ( \frac{v^2}{2}M_{\textrm{br}}^2 + r^2_{\textrm{br}} M^2_{\textrm{br}} \right ) \frac{1}{f} \sin(\frac{\phi_0}{f}) = 0
\end{equation*}
\begin{equation}
    \frac{1}{F} r^2_{\textrm{roll}} \Lambda_H^4 + \frac{v^2}{2f} M^2_{\textrm{br}} = 0
    \label{eq:relaxion_mixing_dvphi}
\end{equation}

\noindent where we use the fact that $\Lambda_H^4 \gg \Lambda_H^2$, $\sin(\frac{\phi_0}{f}) \sim \sin(\frac{\phi_0}{F}) \sim \mathcal{O}(1)$, and $r_{\textrm{br}} < r_{\textrm{br}} M_{\textrm{br}} \lesssim v$. The last inequality is driven by the relation $M_{\textrm{br}} \lesssim v/r_{\textrm{br}}$. 

\noindent Now, let's calculate the double derivatives. Starting with $M_{hh}$ and leveraging Eq.~\ref{eq:relaxion_mixing_dvh}, we have

 \begin{equation*}
    M_{hh} =  \eval{\partial_{h h} V_{\chi \textrm{NBR}}}_{h \rightarrow v, \phi \rightarrow \phi_0 } = \left ( \kappa \Lambda_H^2 - \Lambda_H^2 \cos(\frac{\phi_0}{F}) +  \lambda v^2 - M_{\textrm{br}}^2 \cos(\frac{\phi_0}{f}) \right ) + 2\lambda v^2 ,
\end{equation*}
\begin{equation}
  M _{hh}= 2 \lambda v^2
  \label{eq:M_hh}
\end{equation}

\noindent Continuing on with $M_{h \phi}$, we have

\begin{equation*} \label{eq:dvhphi}
   M_{h \phi} =   \eval{\partial_{h \phi} V_{\chi \textrm{NBR}}}_{h \rightarrow v, \phi \rightarrow \phi_0 } = \frac{v \Lambda_H^2}{F} \sin(\frac{\phi_0}{F}) + \frac{v M_{\textrm{br}}^2}{f} \sin(\frac{\phi_0}{f}),
\end{equation*}
\begin{equation}
    M_{h \phi}\approx \frac{v M_{\textrm{br}}^2}{f} \sin(\frac{\phi_0}{f})
\end{equation}

\noindent on account of the fact that $F \gg f$. We note, also, that $M_{\textrm{h} \phi}$ = $M_{\phi \textrm{h} }$. Lastly, we calculate $M_{\phi \phi}$. We find that  

\begin{equation*}
\begin{split}
    M_{\phi \phi} = \eval{\partial_{\phi \phi} V_{\chi \textrm{NBR}}}_{h \rightarrow v,\, \phi \rightarrow \phi_0}
    &= \frac{v^2 \Lambda_H^2}{2 F^2} \cos\left(\frac{\phi_0}{F}\right)
     + \frac{r_{\textrm{roll}}^2 \Lambda_H^4}{F^2} \cos\left(\frac{\phi_0}{F}\right) \\
    &\quad + \frac{v^2 M_{\textrm{br}}^2}{2 f^2} \cos\left(\frac{\phi_0}{f}\right)
     + \frac{r_{\textrm{br}}^2 M_{\textrm{br}}^4}{f^2} \cos\left(\frac{\phi_0}{f}\right),
\end{split}
\end{equation*}
\begin{equation} \label{eq:dvphiphi}
    M_{\phi \phi} \approx \frac{v^2 M_{\textrm{br}}^2}{2 f^2} \cos(\frac{\phi_0}{f}).
\end{equation}

\noindent When all of the above is put together, the relaxion mass  (Eq.~\ref{eq:relaxion-mass}) can be easily computed.

\subsection{Calculation of Relaxion-Higgs Mixing Angle}
\label{sec:calculation_sin}

\noindent We require that the physical Higgs mass $m_h$ correspond to one of the eigenvalues such that $\det(\mathcal{M}_{h \phi} - m_h^2 \mathbb{1}) = 0$. We find that

\begin{equation}
    M_{hh} = \frac{M_{h \phi}^2}{M_{\phi \phi} - m_h^2} + m_h^2.
\end{equation}

\noindent The eigenvalues of $\mathcal{M}_{h \phi}$ are 

\begin{equation}
    \lambda_h = m_h^2 \qquad \textrm{and} \qquad \lambda_{\phi} = \frac{M_{h \phi}^2}{M_{\phi \phi} - m_h^2} + M_{\phi \phi} = m_{\phi}^2.
\end{equation}

To find the relaxion-higgs mixing angle $\theta$, we recall that once the relaxion-Higgs mass matrix $\mathcal{M}_{h \phi}$ is diagonalized by the matrix $R$, through a similarity transformation which can be considered a rotation, the diagonalized matrix $M_D$ can be obtained. The similarity transformation proceeds as follows:
\begin{equation}
    R^{\intercal} \mathcal{M}_{h \phi} R = M_D.
\end{equation}
This can be rearranged to look like $\mathcal{M}_{h \phi} R = R M_D$. Put explicitly in terms of the matrix components, we have
\begin{align*}
    \pmqty{M_{hh} & M_{h \phi} \\ M_{\phi h} & M_{\phi \phi}} \pmqty{-\sin{\theta} & \cos{\theta} \\ \cos{\theta} & \sin{\theta}} & = \pmqty{-\sin{\theta} & \cos{\theta} \\ \cos{\theta} & \sin{\theta}} \pmqty{m_h^2 & 0 \\ 0 & m_{\phi}^2}, \\
    \pmqty{M_{h \phi} \cos{\theta} - M_{hh} \sin{\theta} & M_{hh} \cos{\theta} + M_{h \phi} \sin{\theta} \\ M_{\phi \phi} \cos{\theta} - M_{\phi h} \sin{\theta} & M_{\phi h} \cos{\theta} + M_{\phi \phi} \sin{\theta}} & = \pmqty{-m_h^2  \sin{\theta} & m_{\phi}^2 \cos{\theta} \\ m_h^2  \cos{\theta} & m_{\phi}^2 \sin{\theta}}.
\end{align*}
We see that we have four equations. One of them yields
\begin{align*}
    M_{h \phi} \cos{\theta} - M_{hh} \sin{\theta} & = -m_h^2 \sin{\theta}, \\
    \qty( m_h^2 - M_{hh} ) \sin{\theta} & = M_{h \phi} \cos{\theta}, \\
    \frac{\sin{\theta}}{\cos{\theta}} & = \frac{M_{h \phi}}{m_h^2 - M_{hh}}, \\
    \tan{\theta} & \approx \frac{M_{h \phi}}{2 m_h^2 + \frac{M_{h \phi}^2}{m_h^2}}, \\
     & \approx \frac{f M_{\textrm{br}}^2 m_h^2 v \sin \left(\frac{\phi_0}{f}\right)}{2 f^2 m_h^4+M_{\textrm{br}}^4 v^2 \sin^2\left(\frac{\phi_0}{f}\right)}, \\
     & \approx \frac{v M_{\textrm{br}}^2}{2 f m_h^2} \sin\left(\frac{\phi_0}{f}\right).
\end{align*}
Assuming that the mixing angle $\theta$ is small we can approximate $\sin{\theta} \approx \tan{\theta}$. Using this approximation, Eq.~\ref{eq:relaxion-mass}, and the assumption that $\sin\left(\frac{\phi_0}{f}\right) \sim \mathcal{O}(1)$, we get 
\begin{equation}
    \sin{\theta} \approx \frac{v M_{\textrm{br}}^2}{2 f m_h^2} \approx \frac{f m_{\phi}^2}{v m_h^2}.
\end{equation}

\acknowledgments 
We would like to thank Pouya Asadi, Spencer Chang, Graham Kribs, and Wolfram Ratzinger for useful discussions and insightful comments. We would also like to thank Wen-Yuan Ai, Shyam Balaji, and Edoardo Vitagliano for pointing us to relevant literature on the strong $CP$ problem and light scalar-Higgs mixing bounds. C.A. acknowledges the support of DOE Award No. DE-SC0011640. J.I. acknowledges the support of NSF Grant No. PHY-2221549.


\bibliographystyle{JHEP}
\bibliography{references}

\end{document}